# Intelligent Edge Computing

Kalgi Gandhi[a], Minal Bhise[b]

[a]Pandit Deendayal Energy University, Gandhinagar, Gujarat, India
[b]Distributed Databases Group, Dhirubhai Ambani University, Gandhinagar, Gujarat, India

ABSTRACT

The number of edge devices in large-scale edge systems is rapidly increasing. Edge devices have limited processing power, memory, and network bandwidth, making resource utilization and data management during edge query processing challenging. Joins are among the costliest database operations in terms of time and resources. The State-of-the-Art edge query processing, Column Imprint-Hash Join CI-HJ, addresses this challenge using equi-height binning to accelerate hash joins. However, it lacks efficiency in real-time processing and scans unnecessary cachelines. This paper presents Workload Aware Column Imprint-Hash Join WACI-HJ, which uses a workload-aware approach to accelerate hash joins. Predicting the upcoming query workload in advance further improves its suitability for real-time edge query processing. WACI-HJ comprises two phases: WACI-HJ Generation Phase, including Pre-processing, Prediction, and Blocking and Hashing modules to compute bins based on the predicted workload before query arrival, and Query Processing and Resource Utilization, which handles query processing and CPU, RAM, and I/O utilization. Evaluations on a benchmark dataset and a real-world Smart Transportation dataset show a 54% reduction in cachelines read and 10% improved query execution time. The proposed technique is effective for both scaled and skewed data. Although PCR is an indirect measure of energy consumption, the work also directly measures energy consumption through energy-efficiency experiments. WACI-HJ shows 1%, 38%, and 49% gain in CPU, RAM, and I/O, respectively. Optimizing cache usage and query execution speeds up real-time traffic analysis, congestion management, and routing in Smart Transportation. Additionally, this technology can be applied to other domains to accelerate edge query processing.

## 1. Introduction

Edge Computing (EC) performs computing at the edge, i.e. the source of the network. In contrast to Cloud Computing, EC brings computation and data storage closer to the data sources [1, 2]. The typical edge system is only a few hops away from the data layer. It enables real-time applications, minimizes data transfer, and facilitates rapid decision-making. Ideal for resource-constrained environments and the Internet of Things (IoT), EC analyzes data locally before transmitting structured data to the cloud.

In the context of the Internet of Things (IoT), if all the large amounts of unstructured or semi-structured data generated by the connected devices are transmitted to the cloud, latency and bandwidth will increase. EC is an intermediate step, as seen in Figure 1. It processes the unstructured or semi-structured data transferred by IoT and sends the processed structured data to the cloud for storage. This data-handling process is crucial for efficient query processing, as it ensures that only the processed data is sent to the cloud, reducing latency and bandwidth usage [3].

EC involves storing, processing, and analyzing data directly on edge devices or nodes rather than sending all data to the cloud. This approach reduces latency by enabling quick query responses and optimizing bandwidth by transmitting only relevant information. It handles and analyzes data closer to its source rather than sending it to distant servers. This minimizes latency, conserves bandwidth, enables real-time analytics, and reduces energy consumption [3].

kalgi_gandhi@dau.ac.in (K. Gandhi); minal_bhise@dau.ac.in (M. Bhise)

### 1.1. Motivation

EC performs computing close to the source of the network, and it is just a few hops away from the data layer. The number of edge-based applications and the associated data are growing rapidly. EC applications in various fields, including Smart Manufacturing [4], Intelligent Transportation Systems [5], Health Monitoring Systems [6], Smart Homes [7], Smart Supply Chain Management [8, 9] and Smart Cities [10] as seen in Figure 2. These devices are actively deployed and utilized in these domains to improve efficiency, safety, and convenience.

Enhancing energy-efficiency and sustainability by locally processing real-time traffic analysis, congestion management, and routing. As a result, EC contributes to a more efficient, responsive, and sustainable smart transportation and logistics infrastructure.

As smart applications continue to develop, the number and variety of connected devices have increased, leading to rapid data growth [3]. From 9.76 billion IoT devices in 2020, the count is projected to reach 34.6 billion by 2030 [11], reflecting widespread adoption across various industries. Consequently, IoT data, which were 44 zettabytes (ZB) in 2020 [12], is expected to reach 246 ZB in 2030 [13]. This explosive increase underscores the crucial role of effective data transmission and storage at the edge to manage and utilize this expanding digital ecosystem effectively.

### 1.2. Problem Statement

The rapid increase in data growth at the edge raises concerns about efficient transmission and storage, given the limited processing power, memory, and network bandwidth of edge devices. Effective data management strategies are essential to handle the expanding data volumes without overloading these devices. Energy-efficient techniques are needed to improve memory usage and minimize data transfer over the network [3].

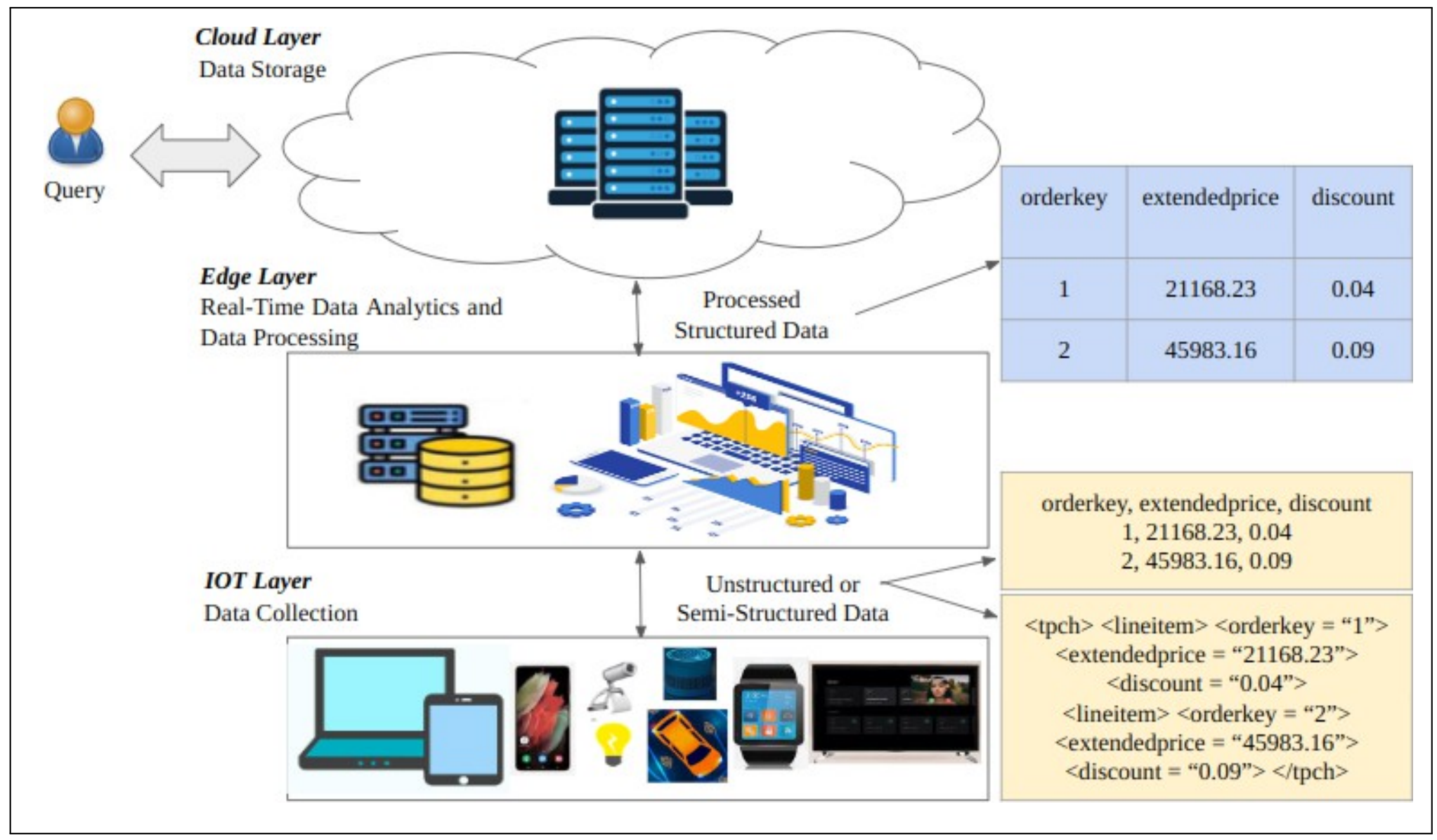

**Figure 1:** Query Processing at Edge

This work proposes a single-edge-node strategy over a decentralized architecture to accelerate query processing, addressing complexity and resource utilization concerns. Joins, particularly critical in resource-constrained edge environments, benefit from the Column Imprint-Hash Join (CI-HJ), which uses equi-height binning for efficient hash joins [14]. The Workload Aware Column Imprint-Hash Join (WACI-HJ) enhances this using a workload-aware approach, precomputing Column Imprint (CI) [15] based on query workload predictions. This approach optimizes performance by preparing data storage ahead of query arrival and monitoring resource usage in edge environments, such as CPU, RAM, and I/O.

### 1.3. Objectives

The main objectives of the work are as follows:

- **To accelerate Query Processing for Edge Systems-** The aim is to accelerate query processing at the edge using WACIs to optimize hash joins and reduce cache misses.
- **To address Real-Time Query Processing for Edge Systems-** The objective is to enable real-time query processing by predicting upcoming workloads by analyzing historical data, trends, and patterns.
- **To monitor Resource Utilization for the Proposed Work-** The objective is to keep track of CPU, RAM, and I/O usage to assess energy-efficiency and effectiveness in resource-constrained edge environments.

### 1.4. Contributions

The main contributions of the work are as follows:

- The Workload-Aware Column Imprint-Hash Join (WACI-HJ) algorithm enhances real-time edge query performance.
- The proposed technique has been demonstrated to work well with scaled and skewed data.
- WACI-HJ is validated using Transaction Processing Performance Council (TPC) TPC-H [16], TPC-D [17], and Metropolitan Atlanta Rapid Transit Authority (MARTA) [18] datasets, demonstrating versatility across diverse domains.
- WACI-HJ outperforms State-of-the-Art Column Imprint-Hash Join (CI-HJ) [14] on key evaluation metrics.
- Reduce latency, improve energy-efficiency, and optimize data access, supporting sustainable EC database management.

The rest of the paper is organized as follows: Section 2 briefly summarizes the Literature Survey. The Proposed Technique is elaborated in section 3, its Implementation Details and Results are discussed in sections 4 and 5, and section 6 concludes the paper.

## 2. Literature Survey

This section discusses the traditional database management and query processing, alongwith edge data management and query processing. Accelerating hash joins is vital to enhance query processing in edge systems. Hence, methods to accelerate hash joins are also elaborated further. Additionally, topics like merging Artificial Intelligence (AI) with EC predicting queries and keeping track of the resources used in EC setups. Additionally, current research issues and open research ideas.

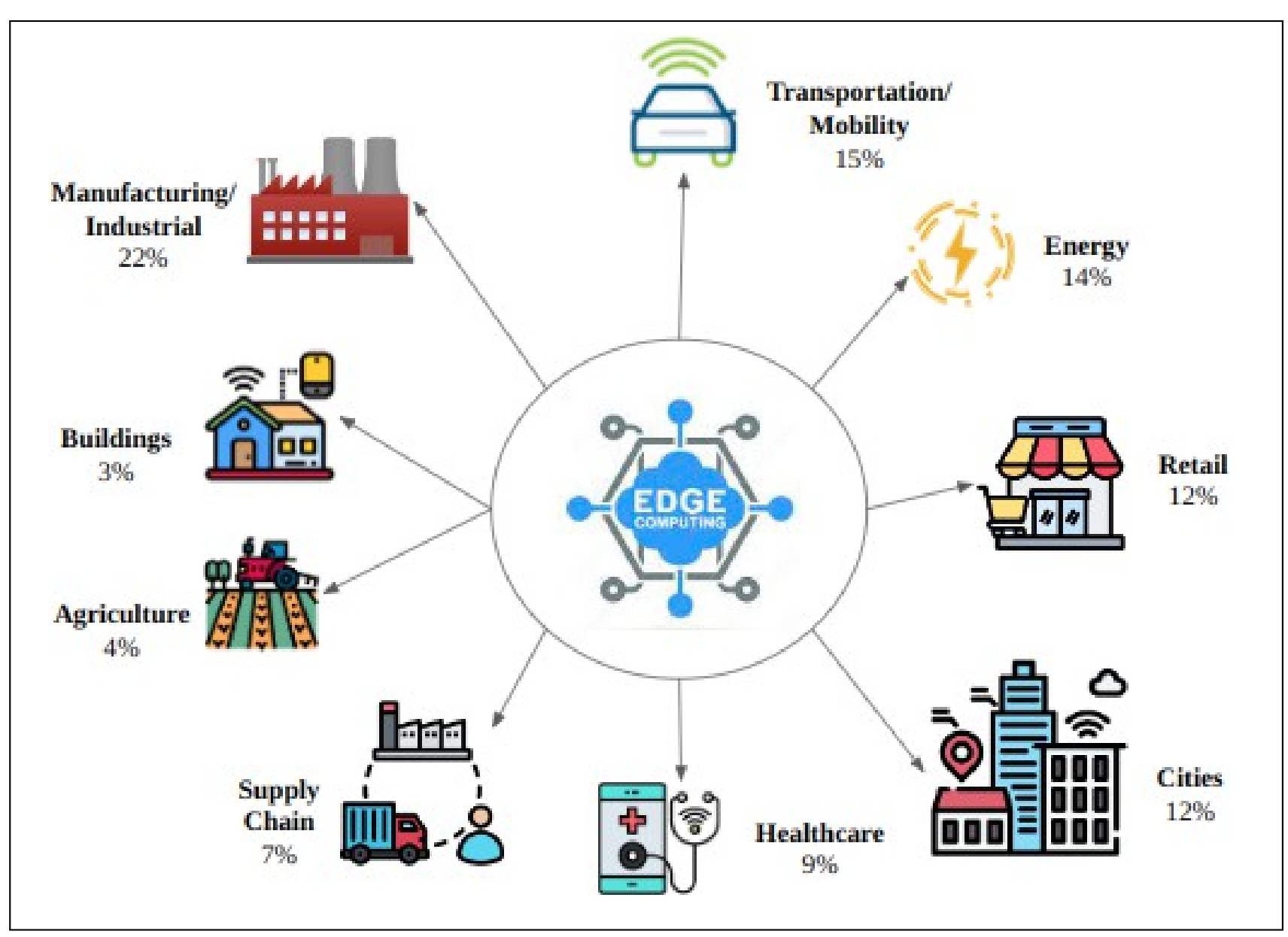

**Figure 2:** Edge Computing Applications

## 2.1. Database Management and Query Processing

Database query processing relies on four main components: the CPU, RAM, cache, and disk storage, each playing a crucial role in executing queries efficiently. The CPU executes queries by parsing, optimizing, and executing them, focusing on generating efficient execution plans. RAM, or main memory, stores data that the CPU can access rapidly, reducing latency compared to secondary storage. Cache, including L1, L2, and sometimes L3 caches, provides quick access to frequently used data, improving query performance through cache hits. Secondary memory, often in the form of HDDs or SSDs, stores persistent data. When data is unavailable in RAM, the CPU retrieves it from secondary storage, although at a slower rate. These components collectively facilitate the efficient execution of database queries.

Database query processing begins with parsing the query to ensure its syntactic correctness and validity, followed by optimization to generate an efficient execution plan. During execution, the CPU, RAM, cache, and disk storage work collaboratively to access and process data, ultimately producing the query result. The optimized execution plan ensures minimal resource usage and execution time. Finally, the query result is presented to the user in a readable format, completing the query processing cycle.

Materialized Views [19] store precomputed query results, reducing computational overhead by eliminating the need to perform complex aggregations or joins repeatedly during query execution. Data Partitioning [20, 21] divides large datasets into smaller partitions, enhancing scalability and improving data access efficiency by allowing queries to target specific partitions rather than scanning the entire dataset. Several researchers have also explored partitioning strategies for different varieties of data [22, 23]. Data Blocking [24] organizes data into contiguous blocks, minimizing disk seeks and improving retrieval speed by storing related data items close together. Data Summarization [25, 26] stores data into compact forms, reducing the data volume and speeding up queries by storing aggregate values instead of individual data points. Indexing structures like Primary and Secondary Indexing [15] provide efficient access paths to data, minimizing disk accesses and optimizing query performance by facilitating rapid data retrieval based on key values. These techniques collectively enhance query processing efficiency in databases, crucial for managing and querying large datasets effectively.

## 2.2. Data Management and Query Processing at Edge

EC research targets latency reduction via efficient communication, resource management, bandwidth optimization, energy-efficient techniques, and robust data security measures to enhance performance and reliability at the edge.

EC transforms data processing by decentralizing computation and storage, reducing latency and enhancing responsiveness. Techniques such as Load Balancing [27] ensure balanced task distribution across edge devices, optimizing performance and preventing overload. Data Compression [28] minimizes data size before transmission, maximizing bandwidth efficiency and accelerating data transfer. Predicate Caching [29] stores frequently accessed data locally, improving query response times, while Predictive Analytics [30] enables proactive decision-making based on historical data insights. Efficient resource management strategies like Data Partitioning [20, 21] and Prefetching [31] enhance processing efficiency and reduce access latency, while techniques such as Parallel Joins [32] enhance scalability. Initiatives in energy-efficient EC, such as Green Cloud Networking (GCN) [33], optimize energy consumption, while robust security measures like Authentication [34] and Encryption [35] ensure data integrity and protect against unauthorized access in decentralized EC environments.

Edge query processing is pivotal for optimizing data management and enhancing responsiveness in EC environments. By leveraging the computational capabilities of edge devices, query processing occurs locally, minimizing data transmission to centralized clouds and reducing latency. However, the limited processing power, memory, and network bandwidth of edge devices pose significant challenges. Techniques such as efficient data encoding for enhanced memory usage and reduced data volume during transmission [36] and leveraging parallelism to split queries into independent sub-queries [37] are crucial for optimizing throughput and latency in EC. Despite such advancements, systems like EdgeReduce [1] lack validation in practical, real-world scenarios and demonstrate limited scalability across large-scale edge environments.

## 2.3. Hash Joins

In resource-constrained edge environments, efficient techniques like Hash Join [38] are crucial. Hash Join is advantageous because it only requires one table to be fully stored in memory while constructing the hash table, making it memory-efficient.

Accelerating hash joins in resource-constrained environments involves several optimization techniques. Data Partitioning, such as the GRACE [39] method, divides large tables into smaller subsets for independent processing, improving performance but sometimes requiring data exchange between partitions, which can delay execution. Prefetching [31] reduces cache misses by fetching data into the cache ahead of time, although it incurs space overhead. Parallel Joins [32] distribute join operations across multiple nodes to enhance performance but can be inefficient with skewed data, leading to node overloads. Memory-conscious techniques like Accelerate Hash Tables (HTA) [40] and no-partitioning [41] dynamically resize hash tables to minimize memory overhead, though this increases complexity.

## 2.4. Edge Intelligence

Integrating AI with EC aims to leverage advanced AI techniques to enhance processing and decision-making capabilities at the network edge. Techniques such as Edge Intelligence (EI) [42, 43, 44, 45, 46, 47] involves techniques such as Regression [30] and Neural Networks [30], including Recurrent Neural Networks (RNNs) [30], handle linear and non-linear data relationships, respectively. ARIMA [48], a time-series forecasting model, is particularly effective for predicting queries with temporal patterns, such as daily or seasonal trends. These models enable real-time data analysis and predictions directly at the data source, reducing latency and bandwidth usage. Despite these advancements, there are significant limitations. Current EI models often lack the appropriate predictive capabilities to handle complex queries, and the resource-constrained edge environments add to the complexity, making robust and comprehensive predictive models challenging to develop.

CI-HJ [14] is an effective strategy for large-scale edge systems that involves pre-computing CIs for every table intended to be joined. By employing these CIs, a hash can be accelerated, reducing the volume of data that must be scanned and compared during the process. The state-of-the-art CI-HJ [14] enhances query performance while minimizing resource utilization at the edge.

CI-HJ is an effective technique to accelerate query processing at the edge, but it has certain limitations due to its data-aware approach. The query leads to unnecessary scans. It lacks the ability to adapt to real-time query processing requirements effectively.

## 2.5. Open Research Issues

The open research issues derived from the literature survey are Latency Reduction, Resource Management, Energy-Efficiency, and AI Integration. Addressing these challenges is crucial for improving edge computing performance and efficiency.

***Latency Reduction***- Current approaches, due to data-aware nature, often lead to unnecessary data scanning and increased processing times, leading to inefficient latency reduction.

***Resource Management***- Effective resource management remains critical for optimizing task allocation and adapting to dynamic network conditions, yet existing solutions lack robust adaptive algorithms.

***Energy-Eficiency***- Insufficient communication protocols and data compression methods are hindering efforts to minimize latency and reduce energy consumption effectively.

***AI Integration***- Integrating AI into EC faces obstacles in predicting query workloads and optimizing real-time data processing, impacting resource utilization and energy-efficiency in resource-constrained environments. Addressing these challenges is essential for advancing EC capabilities and improving overall system performance.

### *2.5.1. Proposed Approach*

The proposed work aims to address the identified gaps in the literature by developing novel solutions for latency reduction, resource management, energy-efficiency, and AI integration, as depicted in Figure 3. Incorporating a workload-aware approach and prediction hash joins can accelerate energy-efficient edge query processing.

***Workload-Aware Approach***- Hash joins for edge query processing can be accelerated by integrating a workload-aware approach. By incorporating workload characteristics and focusing on scanning only relevant data, edge systems can optimize data scanning, minimizing unnecessary data access and accelerating query processing, thereby reducing latency and enhancing overall system efficiency.

***Workload Prediction***- The system optimises computational power and memory utilization by incorporating predictive

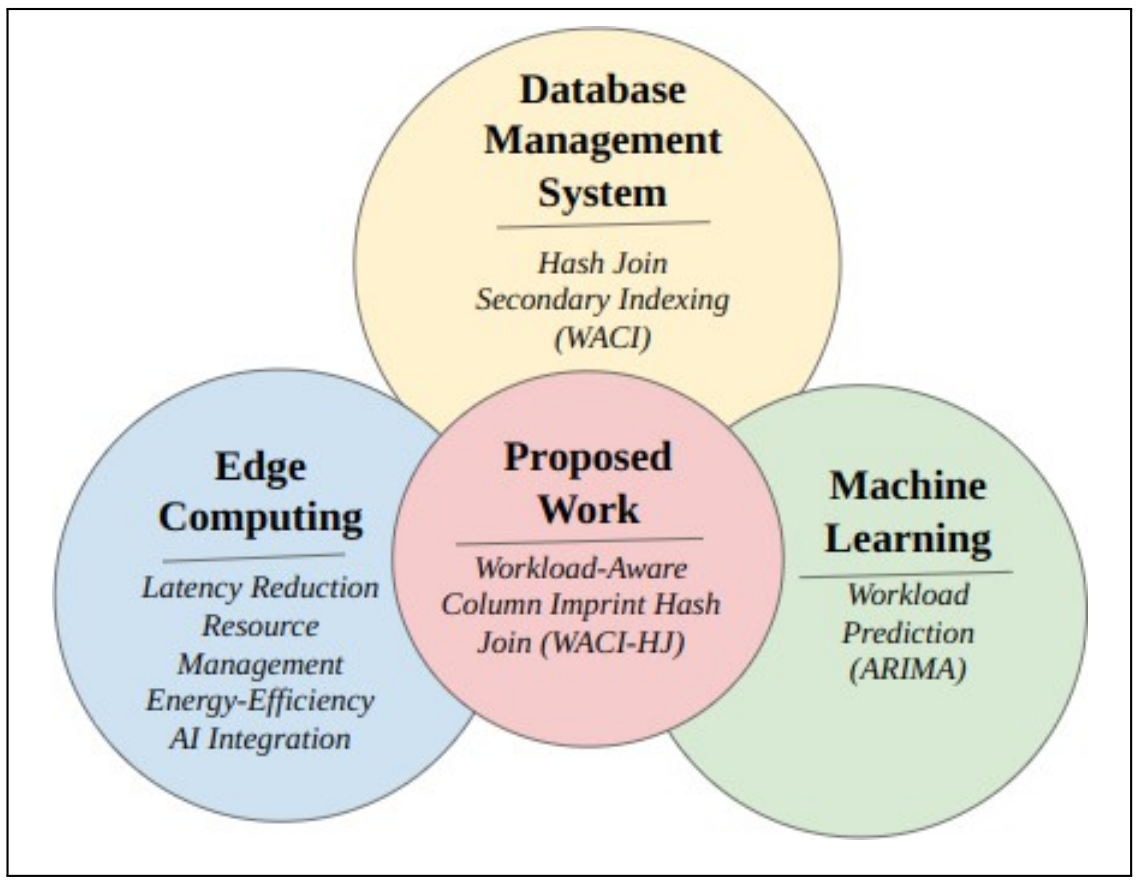

**Figure 3:** Proposed Approach

capabilities to forecast query workloads, ensuring efficient resource management.

***Energy-Eficiency***- Network optimization and energy-efficiency aim to reduce communication latency and conserve energy, further enhancing system performance. Integrating AI and performance monitoring enables the system to predict optimal resource utilization and continuously track performance metrics, ensuring sustained system responsiveness and sustainability.

## 3. Proposed Technique: Workload Aware Column Imprints-Hash Join (WACI-HJ)

Many database systems in cloud environments face scalability and sustainability challenges due to limited bandwidth and high latency [49]. Edge computing addresses these issues by bringing storage and computation closer to the user. Edge computing applications often involve real-time queries, which can be simple (e.g., SELECT queries) or complex (e.g., JOIN queries), requiring more resources and time to process [50]. Given the resource limitations of edge systems, optimizing query execution time, particularly for join operations, is crucial for improving performance. Hash joins effectively enhance performance and scalability, offering significant benefits for handling complex queries in edge systems [38]. CI-HJ [14] is an effective strategy for large-scale edge systems that involves pre-computing CIs for every table intended to be joined. By employing these CIs, a hash can be accelerated, reducing the volume of data that must be scanned and compared during the process. The state-of-the-art CI-HJ [14] enhances query performance while minimizing resource utilization at the edge. CI-HJ was chosen as the primary baseline due to its relevance to columnar processing in resource-constrained edge environments. While techniques like Bloom Filters [51] and Learned Indexes [52] offer optimization benefits, these often involve additional memory overhead, tuning complexity, or assumptions that are less suited for edge workloads.

The proposed work WACI-HJ performs binning according to workload information. Workload patterns and value weights are used instead of a random selection from columns to construct a histogram. The weight is the number of cachelines or the value frequency in the column. The proposed approach applies WACI to accelerate hash joins for large-scale edge systems by pre-computing CIs for each table and incorporating workload information during binning operations. Unlike equi-height binning, binning is now based on query requirements and frequencies. In WACI-HJ, binning relies on workload information from the building column, followed by imprint vector generation for each cacheline of the probe column. This optimized technique aims to reduce cache misses during the hash join operation. WACI-HJ with workload prediction, addresses the challenges of real-time query processing in edge systems. By analyzing historical data trends and patterns, the forecaster within WACI-HJ predicts upcoming workloads, enabling the system to allocate resources and optimize query execution.

### 3.1. WACI-HJ Storage and System Architecture

Storage and architecture are crucial for effectively organizing and preserving data reliability. WACI-HJ Storage and Architecture is specifically tailored for resource-constrained edge applications. It features a compact, lightweight design optimized for limited computing resources.

#### *3.1.1. WACI-HJ Storage Architecture*

Figure 4 shows the storage architecture for WACI-HJ. Data, queries, and the query workload are passed as inputs to the edge phase for processing. The data is then hashed. Data blocks are generated using the WACI approach. Further, when a query arrives, it must scan only these WACIs to find the relevant tuples. Additionally, the utilization of resources, such as CPU, RAM, and I/O, is monitored to check the algorithm energy-efficiency.

#### *3.1.2. WACI-HJ System Architecture*

WACI-HJ System Architecture is shown in Figure 5. It optimizes hash join binning by integrating workload to enhance cache usage and query processing efficiency. It predicts future workloads to allocate resources effectively for improved system performance. WACI-HJ comprises of two main phases: Phase I WACI-HJ Generation and Phase II Query Processing and Resource Utilization. The WACI-HJ Generation phase includes three modules: Pre-processing, Prediction, and Blocking and Hashing. In this phase, bins are computed based on the predicted workload before the query arrives, enhancing algorithm performance. The Query Processing and Resource Utilization phase comprises two modules: Query Processing and Resource Utilization, where queries are processed and CPU, RAM, and I/O utilization are monitored.

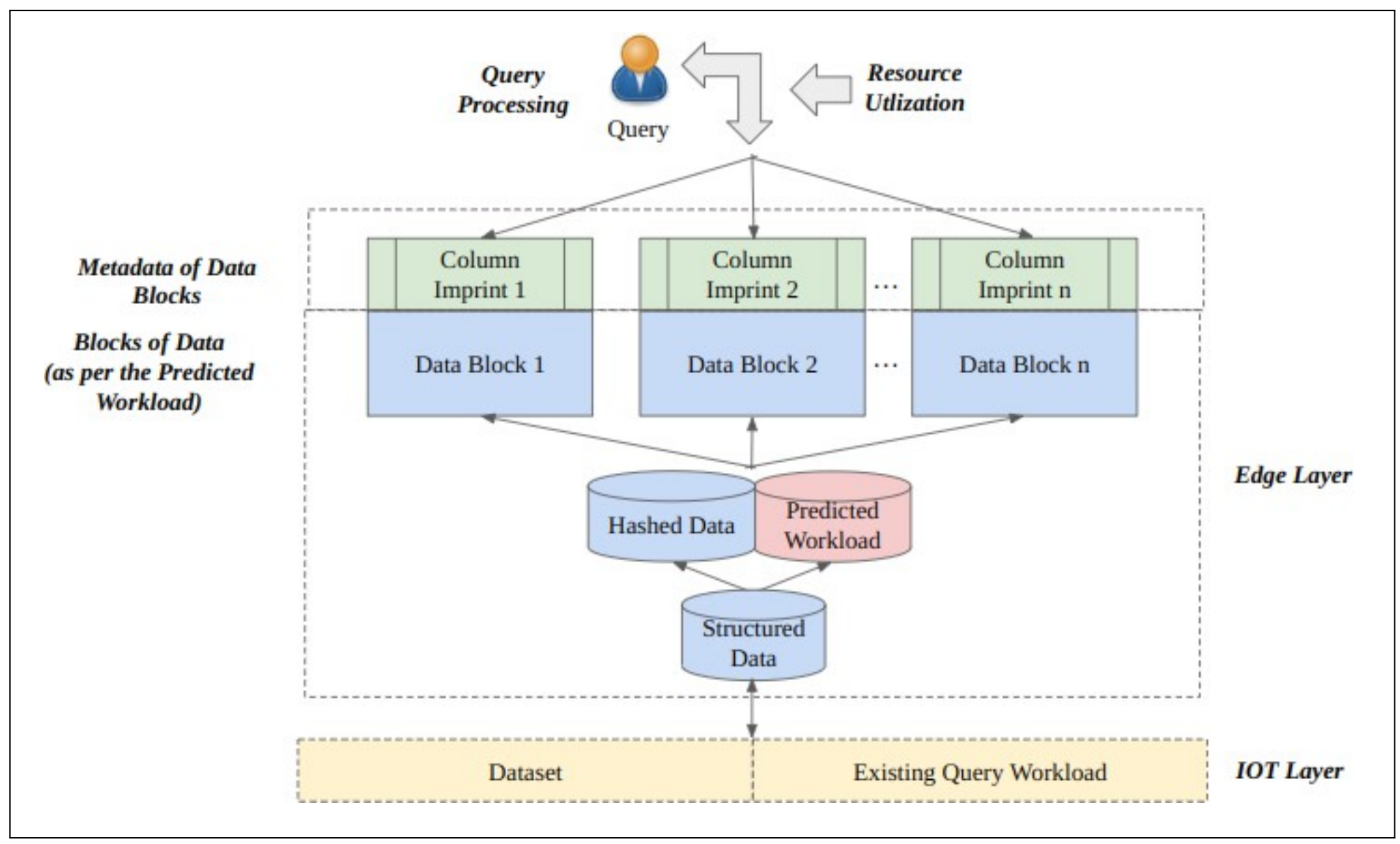

**Figure 4:** WACI-HJ Storage Architecture

## 3.2. WACI-HJ Phases

WACI-HJ uses a workload-aware approach to accelerate hash joins. Additionally, Workload Prediction is used to predict the upcoming query workload. WACI-HJ comprises of two main phases: Phase I WACI-HJ Generation and Phase II Query Processing and Resource Utilization, further discussed in this section.

**Phase I: WACI-HJ Generation**

WACI-HJ Generation Phase comprises of three modules: the Pre-processing module prepares the dataset, the Prediction Module forecasts future query ranges, and the Blocking and Hashing Module organizes data for efficient query processing.

***Module 1: Pre-processing***- Module 1 is the Pre-processing module. It loads the dataset and queryset, generates a workload based on Zipf's Law [53] to reflect real-world query frequency, and creates a Usage Matrix to identify commonly used attributes and its data ranges.

***Module 2: Prediction***- Module 2 is the Prediction module. It clusters data ranges applying the ARIMA [48] model to past data to predict future query frequencies for each cluster and allocates bins to clusters based on the predicted workload.

***Module 3: Blocking and Hashing***- Module 3 Blocking and Hashing defines the bin borders, creates data blocks according to the bins assigned to each cluster, and hash and store data using WACIs, generating WACI-HJ as the output.

**Phase II: Query Processing and Resource Utilization**

In the WACI-HJ framework, Phase II has been added. During this phase, a query is executed on the WACI-HJ output, and resource utilization is monitored.

***Module 4: Query Processing***- Module 4 is the Query Processing module. The system updates the query workload and data blocks if required. If workload prediction is needed, WACI-HJs are generated using the predicted workload. Otherwise, WACI-HJ is generated using the existing query workload. The generated WACI-HJs are stored in secondary memory, and only required cachelines are fetched to the main memory and scanned to generate query output.

***Module 5: Resource Utilization***- Further, in Module 5 the Resource Utilization module, CPU, RAM, and I/O usage are monitored during WACI-HJ query execution. WACI-HJ optimizes CPU utilization by reducing processed cachelines, improving RAM utilization via optimized data access patterns, and enhancing I/O efficiency through predictive prefetching to boost system performance.

## 3.3. WACI-HJ Data Structures and Algorithm

WACI-HJ comprises of two phases: Phase I WACI-HJ Generation and Phase II Query Processing and Resource Utilization. During Phase I, the bins are computed based on the workload before the query arrives, enhancing the algorithm performance. The query is processed during Phase II, and CPU, RAM, and I/O utilization are monitored.

### *3.3.1. Algorithm 1: WACI-HJ Generation*

WACI-HJ data structures are shown in Table 1. Data Structure used for Module 1, Usage Matrix Basket (UMB) lists queries and its attributes. For Module 2, the Query Workload List (QWL) holds the total query frequency count. The Cluster-Range Table (CRT) comprises of ranges. The Cluster-Frequency Table (CFT) contains the query range and weekly arrival frequency.

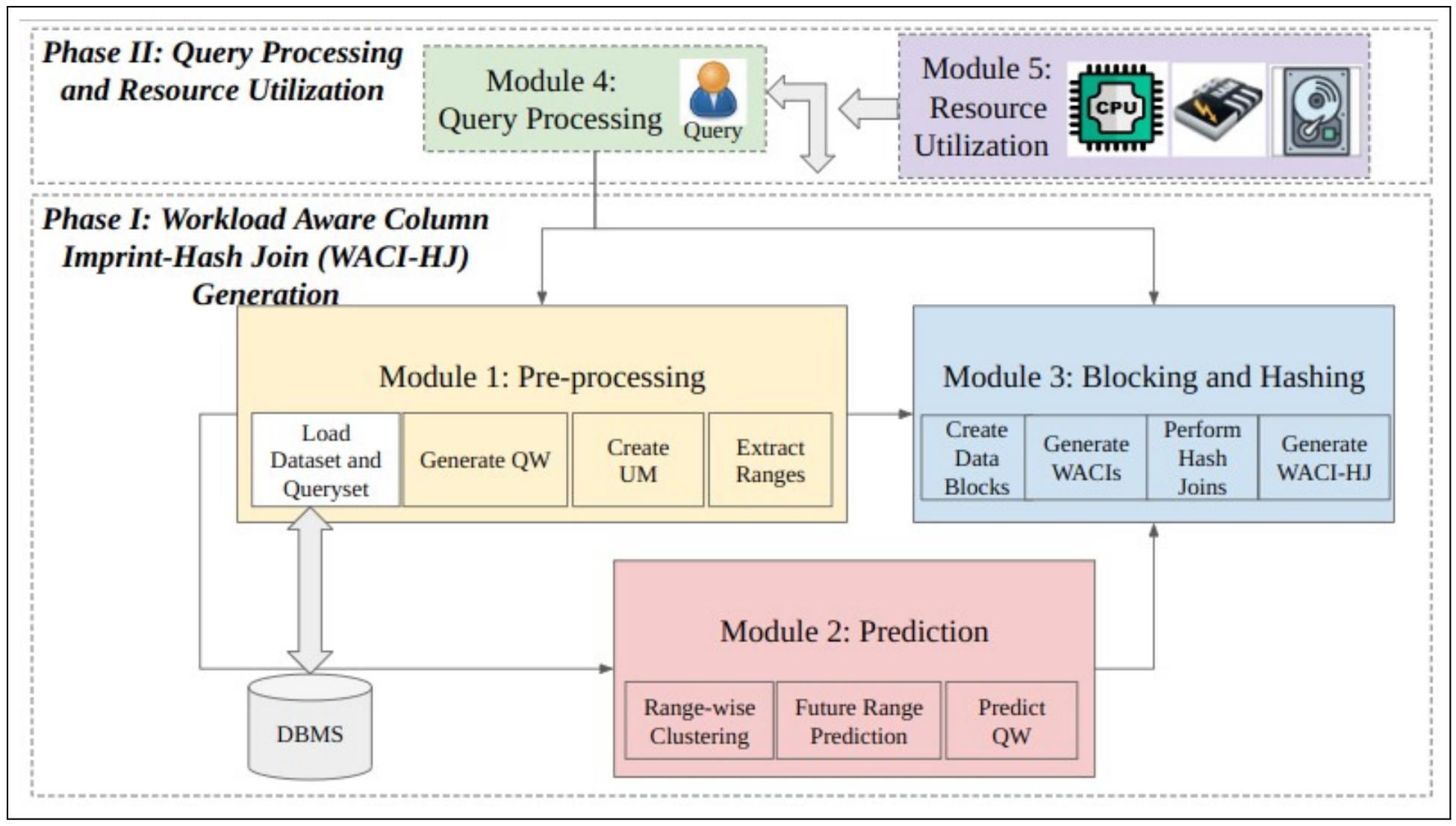

**Figure 5:** WACI-HJ System Architecture

WACI-HJ algorithm is splitted into, WACI-HJ Generation Algorithm, Query Processing Algorithm, and Resource Utilization Algorithm.

The WACI-HJ Generation Algorithm constructs a workload-aware hash table optimized for edge query processing by predicting the query workload beforehand query execution. It takes as input tuples Ti and queries Qj, and operates in three modules: *Pre-processing, Prediction, and WACI-HJ Blocking and Hashing*. The Pre-processing module analyses historical queries to extract frequently accessed attributes and build a Usage Matrix. The Prediction module applies the ARIMA model to forecast future workloads, generating a Predicted Query Workload List. Based on this, the Blocking and Hashing module dynamically assigns bins and computes bin borders to group relevant value ranges. These bins guide the creation of imprint vectors, highlighting relevant cachelines and eliminating unnecessary scans. The final output is a hash table constructed using only tuples within predicted bins, enabling efficient probing and reducing CPU, RAM, and I/O overhead in edge environments.

The time and space complexities of WACI-HJ are $O(2^x(n * t) * 64)$ and $O(n)$ respectively, where x is the log of the number of bins, n represents the total number of tuples in the probe column, i.e. no_of_cachelines*64, and t is the size of the data type.

### 3.3.2. *Algorithm 2: Query Processing*

The Query Processing Algorithm uses data structures such as QueryVector which represents the imprint vector of the incoming query, and the WACI-HJ Generation Output, which contains the imprint vectors generated by the WACI-HJ algorithm.

The Query Processing Algorithm runs incoming queries using the WACI-HJ table. When a Qj query arrives, it creates a QueryVector to show the attributes the query needs. Then it checks if workload prediction is enabled. If so, it loads the WACI-HJ structure built from the Predicted Query Workload List. Further, the algorithm scans the WACI-HJ Output to find relevant cachelines, using the imprint vectors and bin mappings. It performs a bitwise OR between the QueryVector and the matching imprint vectors to find overlaps. This selective filtering produces the Query Output while avoiding unnecessary scans, reducing cacheline access, and making query processing more efficient.

The time and space complexities of Query Processing are $O(n * f)$ and $O(n * m)$ respectively, where n represents the total number of tuples in the probe column, i.e. no_of_cachelines*64, m is the number of queries, and f is the query frequency count.

### 3.3.3. *Algorithm 3: Resource Utilization*

For Module 5, Resource Utilization involves CPU and RAM utilization during query processing, as well as I/O efficiency.

The Resource Utilization Algorithm is a key tool for system administrators and developers to oversee algorithm performance and identify potential bottlenecks or issues. Upon initiation, the algorithm first configures essential parameters for the monitoring process. It then executes the specified monitoring command, leveraging external tools tailored to the type of resources to be monitored, whether CPU and RAM utilization or I/O operations. Subsequently, the algorithm captures the output generated by the monitoring command, storing it within the RM Output variable for further processing.

The time and space complexities of Resource Utilization are $O(n * f)$ and $O(n * m)$ respectively, where n represents the total number of tuples in the probe column, i.e. no_of_cachelines*64, m is the number of queries, and f is the query frequency count.

## 3.4. Integrating WACI-HJ Phases

WACI-HJ comprises of two phases: Phase I WACI-HJ Generation includes Pre-processing, Prediction, and Blocking and Hashing modules to compute bins ahead of query arrival, and Phase II Query Processing and Resource Utilization handles query processing and monitoring of CPU, RAM, and I/O utilization. During Phase I, Module 1 initializes by loading the benchmark dataset and queryset, generating queries using Zipf's Law [53], and constructing a Usage Matrix to identify common attribute usage. This information helps in clustering data based on attribute relevance. In Module 2, cluster usage frequency over weeks is computed, and the ARIMA technique [48] predicts future usage, facilitating bin assignment to clusters. In Module 3, bins are assigned, borders are set, and data is divided into blocks according to predicted cluster usage, storing them as WACIs.

**Table 1**
WACI-HJ Data Structures

| Qj | Aj |
|---|---|
| Q1 | route_id, trip_id, trip_headsign, departure_time, actual_departure_time, delay |
| Q2 | route_id, route_short_name |
| Q3 | route_id, trip_id, trip_headsign, departure_time, actual_departure_time, delay |

(a) for Module 1: UMB

| Qj | QFj |
|---|---|
| Q1 | 50 |
| Q2 | 25 |
| Q3 | 17 |

(b) for Module 2: QWL

| Ci | CRi |
|---|---|
| C1 | 1-1.5M |
| C2 | 1.5M-3M |

(c) for Module 2: CRT

| Ci | QFj | Week |
|---|---|---|
| C1 | 4 | 1 |
| C2 | 17 | 3 |

(d) for Module 2: CFT

| QueryVector |
|---|
| 1000 |

(e) for Module 4: QueryVector

| Qj | QFj |
|---|---|
| 1 | 1000 |
| 2 | 0001 |

(f) for Module 4: WACI-HJ Generation Output

| Query/Resources | CPU Utilization (%) | RAM Utilization (%) | I/O Efficiency (KB) |
|---|---|---|---|
| Q1 | 1.0 | 1.2 | 0 |
| Q2 | 31.7 | 1.4 | 26480 |
| Q3 | 53.8 | 1.4 | 3496 |

(g) for Module 5: Resource Utilization

Furthermore in Phase II, Module 4 upon query arrival, the workload and WACI-HJ Output are updated if required. WACI-HJs, stored in secondary memory, bring the necessary cachelines to the main memory for query output. Further in Module 5, CPU, RAM, and I/O utilization are monitored during query execution.

WACI-HJ time and space complexities are $O(nf.2^{x}(nt)64)$ and $O(n + nm)$ respectively, where n represents the total number of tuples in the probe column, i.e. no_of_cachelines*64, f denotes query frequency, x is the logarithm of bin count, and t signifies data type size. WACI-HJ complexity is a little higher than CI-HJ, due to its workload-awareness and prediction capabilities. WACI-HJ is an energy-efficient algorithm that provides optimized data storage and efficient real-time query processing for resource-constraint edge environments.

## 3.5. Technical Challenges

Implementing the WACI-HJ algorithm presents several technical challenges. Choosing an appropriate workload prediction technique was crucial, and ARIMA was selected for its robustness and accuracy. Integrating hash joins with WACI required careful allocation of bins based on workload information to improve efficiency. Identifying cachelines of interest involved optimizing the scanning process. Additionally, integrating Resource Utilization tools like top and iotop with WACI-HJ was essential to assess system energy efficiency and monitor CPU, RAM, and I/O usage.

***Choosing Appropriate Workload Prediction Technique***- Selecting a suitable prediction technique for Module 2 posed challenges due to varying factors such as query complexity, data distribution and dataset size. We adopted the ARIMA model [48] for workload prediction because of its robustness in modelling time series data and its proven accuracy and reliability in forecasting system behaviour, as discussed in Section 2.4. ARIMA offers a favourable trade-off between computational efficiency and predictive performance. In contrast to more resource-intensive techniques, such as LSTM or hybrid models such as Prophet, ARIMA is lightweight, interpretable, and better suited for deployment in resource-constrained edge environments [54].

***Integrating Hash Joins with WACI***- In Module 3, a build and probe column is used in a hash join. The smaller column is considered a build column to create a hash table. Integrating WACI with hash join was challenging as the bins are allocated based on workload information, and bin borders are calculated and stored the hashed data for improved efficiency.

***Identifying Cachelines of Interest***- In Module 4, the visit list algorithm [14] optimizes scanning imprint vectors in the probe column by utilizing each bin. The hash table is scanned bin-wise during

**Algorithm 1** Workload Aware Column Imprint-Hash Join (WACI-HJ) Generation Algorithm

```
Input: Tuples T_i, Queries Q_j
Output: WACI-HJ Output
//Module 1: Pre-processing
Initialize UMB
for query Q_j do
   A_j = Extract_Attributes_From_Query(Q_j)
   for attribute A_i in A_j do
      if A_i exists in UMB then
         Add_Query_Index(A_i, Index_of(Q_j), UMB)
      else
         Set_Query_Index(A_i, [Index_of(Q_j)], UMB)
      end if
   end for
end for
//Module 2: Prediction
Create CRT C_i and CR_i
CFT = number of weeks * CRT
Cluster_week_freq = Σ^{no_of_queries} QF_j
CF_i = array(cluster_week_freq)
model = auto_arima(CF_i, suppress_warnings = True)
PredictedQWL = model.predict(n_periods = 1, return_conf_int = True)
Bin_Allocation = allocate_bins(PredictedQWL)
bin_borders = compute_bin_borders(Bin_Allocation)
//Module 3: Blocking and Hashing
imp_vec = no_of_cachelines × sizeof(ImprintVector) for i = 0 to no_of_cachelines - 1 do
   imp_vec[i] = no_of_bins × sizeof(datatype)
end for
for j = 0 to no_of_cachelines - 1 do
  for val = 1 to cachelineSize do
      bin = getBin(val, bin_borders)
      if bin ∈ PredictedQWL.keys() then
         imp_vec[j][bin] = 1
      end if
   end for
end for
for val = 0 to build_col_size - 1 do
   if getBin(col[val], bin_borders) ∈ PredictedQWL.keys()
then
      HashArray[hashFunction(col[val])] = val[i]
   end if
end for
for i = minimum_bin_required to maximum_bin_required do
   while list_temp[i] do
      q = list_temp[i] → data
      for w = q × cacheline_size to min(q × (cachelines_size + 1), size_probe_col) do
         if probe_col[w] ≤ bin_borders[i] or probe_col[w] ≥ bin_borders[i + 1] then
            continue
         end if
         if HashArray[hashFunction(probe_col[w])] ≠ -1 then
            store BAT Values corresponding to both columns in WACI-HJ Output
         end if
      end for
      list_temp[i] = list_temp[i] → nx
end while end for
```

**Algorithm 2** Query Processing Algorithm

```
Input: Q_j
Output: Query Output
//Module 4: Query Processing
Generate QueryVector Q_j
if prediction is used then
   Generate WACI-HJ using PredictedQWL
else
   Generate WACI-HJ using existing QF_j
end if
if Q_j is present in the queryset then
   Update QF_j
else
   Update QF_j and B_i
end if
for i = 0 to no_of_cachelines - 1 do
 if ∃ bin ∈ PredictedQWL.keys() such that imp_vec[i][bin] = 1 then
     for val = 1 to cachelineSize do
      bin = getBin(val, bin_borders)
         if HashArray[hashFunction(val)] ≠ -1 then Store BAT Values in Query Output
         end if
      end for
   end if
end for
```

**Algorithm 3** Resource Utilization Algorithm

```
Input: Result Path, RMType, RMCommand
Output: RM Output
//Module 5: Resource Utilization
Set Parameters
Execute RMCommand using External Tools
Read Resource Utilization Output into RM_Output
while RM_Output ≠ NULL do
   for each line in RM_Output do
      Filter Resource Utilization Data for Q_j
      Write Utilized Data to Result Path
   end for
end while
```

probing, and comparisons are made with cachelines containing data falling within that bin. Matches trigger join operations, allowing efficient scanning of the entire hash table and cachelines to perform join operations for matched values.

***Integration of Resource Utilization Tools with WACI-HJ***- In Module 5, tools like top and iotop monitor and analyze resource utilization, offering textual output. This format facilitates easy filtering and faster storage in CSV format, simplifying the analysis. Integrating tools like top and iotop with WACI-HJ enables us to assess the system energy-efficiency directly. This integration monitors CPU, RAM, and I/O usage, providing insights into resource utilization and system performance.

## 4. Implementation Details

This section provides a comprehensive overview of the hard-ware and software setup, dataset and queryset specifications, and evaluation parameters used for the implementation. It outlines the technical environment and resources utilized to demonstrate the effectiveness of the WACI-HJ algorithm.

### 4.1. Hardware and Software Setup

The machine hardware configuration included a quad-core Intel i3-2100 CPU clocked at 3.10 GHz. The system has 32 GB

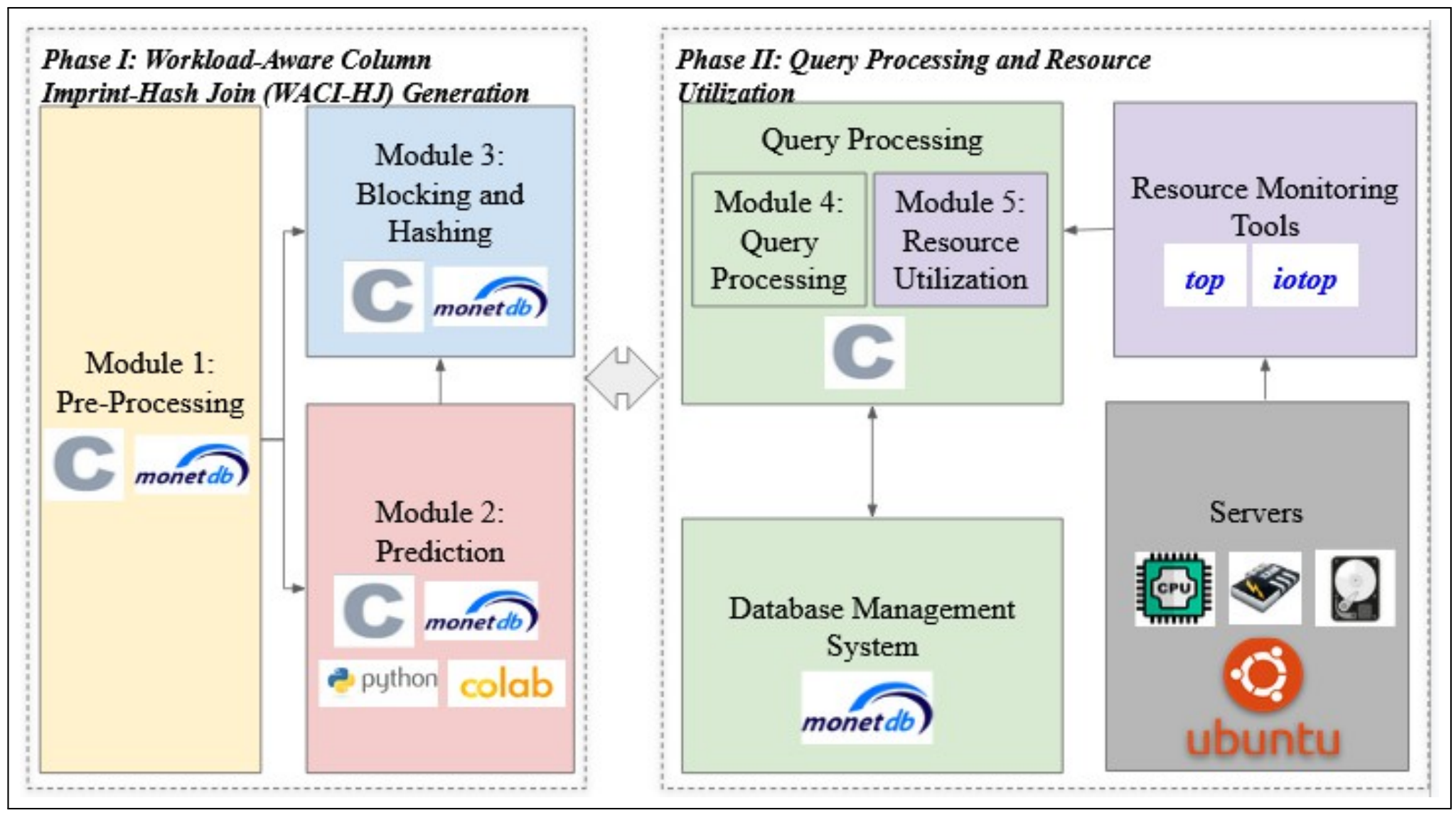

**Figure 6:** Implementation Block Diagram

of RAM, temporarily storing data that the CPU needs to access quickly. It has 500 GB of hard disk space for storing data and files. Additionally, the system includes hardware with L1, L2, and L3 caches sized at 64 Kilobyte (KB), 512 KB, and 3 MB respectively, enhancing processing speed by storing frequently accessed data close to the CPU.

The software required to implement the WACI-HJ is the VS Code IDE 1.83.1 [55] for C language development, and the MonetDB 11.23.3 [56], an open-source column store database for implementing CIs. Google Colab [57] for Python 3.10, supporting machine learning computing, is used for prediction tasks.

The implementation block diagram is shown in Figure 6. The initial block includes the first three modules of the WACI-HJ Generation Phase: (1) Pre-processing (2) Prediction (3) Blocking and Hashing. The Pre-processing module and WACI-HJ Generation module are implemented using the C language for programming purposes, as well as MonetDB, which is a column-oriented database. When workload prediction is induced with WACI-HJ, the prediction code is scripted using Python on the Google Colab [57] platform. The versatility and libraries of Python make it ideal for developing prediction algorithms. Google Colab provides an environment that offers powerful computing resources, enabling the efficient execution of complex prediction tasks.

The Query Processing and Resource Utilization Phase is shown in the next block. The query reads and writes the relevant data from the database per the query request.

The real-time resource utilization data is monitored using tools like top [58] and iotop [59]. These tools monitor the total CPU, RAM, and I/O hardware resource utilization of entire systems and individual processes. Utilization of CPU and RAM is monitored using the top tool, while I/O efficiency is monitored using the iotop tool. The entire system is implemented using the Ubuntu 20.04.6 Long-Term Support (LTS) operating system.

## 4.2. Dataset and Queryset

The implementation of WACI-HJ is demonstrated using a benchmark dataset, a standardized dataset used to evaluate algorithms. These are usually well-documented, public, and widely used in research. Alongwith, real-world datasets contain authentic data from sources like sensors.

***TPC-H [16]***- TPC-H is a Benchmark dataset from the Supply Chain domain. It is a decision-support dataset. It simulates a data warehousing scenario and includes eight tables: customer, orders, lineitem, part, supplier, partsupp, nation, and region. The orderkey attribute is used for experimentation purposes, as it is the most frequently used attribute. Here, here l_orderkey is joined with o_orderkey. TPC-H is a uniform dataset with a 1 GB data size. To test the algorithm for scaled data, it is scaled upto 10x. To demonstrate WACI-HJ, range queries with joins are required. The TPC-H dataset is business-oriented. Although it consists of 22 OLAP queries, only 9 contain joins with orderkey. Additionally, 11 range queries with various numbers of joins were designed and added to the original queryset to make it comprehensive. So, a total of 20 queries were used for the demonstration.

***TPC-D [17]***- Most earlier work on edge systems tested its algorithms on uniform data. However, in the real-world, the dataset is not always uniformly distributed. So, it becomes crucial to evaluate the algorithm for a skewed dataset. The benchmark TPC-D [17] dataset is used to demonstrate WACI-HJ. It is a skewed version of the TPC-H dataset. WACI-HJ is evaluated at different levels of data skewness. It is evaluated at 60%, 70%, 80%, 90%, and 99%. 50% is uniformly distributed. TPC-D is a skewed version of TPC-H. Hence, these two have a common queryset.

***MARTA [18]***- To demonstrate the proposed technique on different datasets in addition to TPC-H [16] and TPC-D [17], MARTA [18] is used. Real-world dataset MARTA is related to Smart Transportation, with one of the highest number of EC applications. MARTA [18] comprises of 13 tables: agency, apcdata_NorthAve, avl_otpdata_week, calendar, calendar_dates, parking_info, parking_stopinfo, routes, shapes, station, stop_times, stops, trips. For experimentation purposes, the route_id attribute is used. MARTA dataset is of 93.1 MB in size. It is scaled upto 10x. To showcase WACI-HJ effectiveness, 10 customized range queries with various numbers of joins are included for demonstration purposes.

### 4.3. Set of Experiments

The experiments are categorized into various sets to evaluate the robustness and performance of the algorithm. Initially, Set 0 configures the system by determining the optimal bit size and number of clusters. Set I performs basic experiments on the default data size, while Set II scales the data size from the default to 10 times larger. Set III tests varying skewness levels from 50% to 99%. Finally, Set IV focuses on energy-efficiency experiments, monitoring CPU and RAM utilization, and I/O efficiency.

***Set 0: Configuring the System***- In the beginning, Set 0 sets the Optimal Bit Size and Optimal Number of Clusters to maximize the algorithm effectiveness. After setting the optimal parameters for the experiments, the experiments are categorized into four sets.

***Set I: Basic Experiments***- Set I is the Basic Experiments performed on the standard dataset size of x. The TPC-H dataset default data size is 1 GB, while the MARTA dataset is 0.97 GB.

***Set II: Data Scaling Experiments***- Set II is for the Data Scaling Experiments performed to evaluate how these techniques perform with different data sizes. Here, the dataset size varies from x to 10x.

***Set III: Varying Skewness Levels Experiments***- Most earlier work on edge systems tested its algorithm on uniform data. However, the data generated in practice is skewed [60]. So, evaluating the WACI-HJ algorithm on a skewed dataset becomes crucial for checking the robustness of the algorithm. Set III is of Varying Skewness Levels, where the skewness level is varied from 50% (uniformly skewed) to 60%, 70%, 80%, 90%, and 99% of skewed data.

***Set IV: Energy-Eficiency Experiments***- Finally, the PCR parameter is an indirect parameter to measure the energy-efficiency of WACI-HJ. However, to measure it directly, Set IV Energy-Efficiency Experiments were performed, which monitored CPU and RAM utilization and I/O efficiency. These experiments monitor the resource utilization of the query execution, aiming to assess the energy-efficiency of the system when handling different types of data and workloads.

### 4.4. Evaluation Parameters

The evaluation parameters are divided into Input, Output, and Resource Utilization categories. Input Parameters include Bit Size 8 to 128 bits, Data Size scaled from x to 10x, and Skewness Levels 50% to 99%. Output Parameters measure system performance through Percentage of Cachelines Read (PCR), Query Execution Time (QET), and Algorithm Execution Time (AET). Resource Utilization Parameters assess CPU and RAM utilization, as well as I/O efficiency, providing a comprehensive evaluation of the efficiency and performance of the algorithm.

***Input Parameters***- Input Parameters specify the variables and configurations applied to the system during experiments. Input Parameters are Bit Size ranging from 8 to 128 bits, Data Size scaled from x to 10x of a base 1 GB for TPC-H and 0.97 for MARTA, and Skewness Levels examined at 50% (uniformly skewed), 60%, 70%, 80%, 90%, and 99% for assessing data distribution uniformity.

***Output Parameters***- Output Parameters quantify the system performance and outcomes. Output Parameters are PCR indicates the proportion of blocks accessed by a query compared to the total number of blocks, QET measures the time taken for each query to execute, typically in seconds, and AET denotes the overall duration of the algorithm execution, also measured in seconds.

***Resource Utilization Parameters***- Resource Utilization Parameters including CPU Utilization, RAM Utilization, and I/O Efficiency, assess the computational resources utilized during system operations. These parameters collectively form the basis for evaluating the efficiency and performance of the WACI-HJ system under various experimental conditions.

## 5. Results and Discussions

This section presents results for indexing techniques such as the state-of-the-art technique CI-HJ [14] and the proposed work WACI-HJ, with and without Workload Prediction, are discussed in this section. These techniques are evaluated based on predefined input, output, and Resource Utilization parameters. The results are derived for benchmark TPC-H [16], TPC-D [17], and the skewed version of TPC-H. Alongwith, a real-world dataset, MARTA [18], which belongs to the Smart Transportation domain. The Smart Transportation domain has the highest growth in edge data applications. All the results presented in this section are averaged over all the queries of the queryset.

Basic Experiments, Data Scaling Experiments, and Energy-Efficiency Experiments using the MARTA [18] dataset are discussed in this section. However, due to the unavailability of the skewed version of MARTA, the experiments related to Varying Skewness Levels were omitted for this dataset.

### 5.1. Set 0: Configuring the System

Configuring a system involves setting up optimal parameters. The proposed system bit size and the number of clusters are set at optimal values for better query performance.

**Setting Optimal Bit Size**

The experiments were performed for various CI bit sizes. Bit sizes vary from 8-bits to 128-bits. Parameters like PCR and QET are the best set. The PCR for WACI-HJ is significantly less at 32-bits, i.e. 2% to 9%, compared to other bit sizes, as shown in Figure 7. This is optimal compared to other bit sizes, as it uses fewer resources and leads to fewer cache misses. In edge systems, optimizing resource utilization is very crucial.

In Figure 8, the QET for WACI-HJ is minimal when considering a size of 32-bits, i.e. 2% to 5% faster than other bit sizes. This leads to a reduction in latency.

**Setting Optimal Number of Clusters**

As shown in Figure 9, the MARTA dataset [18] achieves better performance when the number of clusters is set to 2. In this setting, the system scans approximately 1% fewer cachelines compared to configurations with 4 or 6 clusters. This improvement occurs because MARTA exhibits optimal behaviour when the number of clusters aligns closely with the range of data query access.

Furthermore, as shown in Figure 10, the QET for MARTA also improves when using 2 clusters, having a performance gain of 1% to 3% over 4 or 6 clusters. This gain is attributed to the reduced number of cachelines scanned, which lowers memory access latency and improves overall execution efficiency.

Consequently, 32-bit size is used consistently across all experiments, as the bit size is system dependent hence, it remains fixed for all datasets. This configuration results in improved L3 cache utilization and reduced I/O overhead. The number of clusters, depends on the dataset size, is set to 4 for TPC-H and TPC-D, and 2 for MARTA.

### 5.2. Set I: Basic Experiments

The default data size for the MARTA [18] dataset is 0.97 GB. The results of PCR and QET for the basic experiments are depicted in this section.

**Percentage of Cachelines Read**

In the MARTA dataset, queries are classified based on selectivity and frequency to capture distinct access patterns. Selectivity is used to divide queries into Low Selectivity (LS), returning less than 0.6% of the column, and High Selectivity (HS), returning more than 0.6%. These are further categorized by frequency using a threshold of 15, resulting in four subgroups: HF-HS (High Frequency–High Selectivity), HF-LS (High Frequency–Low Selectivity), LF-HS (Low Frequency–High Selectivity), and LF-LS (Low Frequency–Low Selectivity). However, for TPC-H selectivity is set to 0.8 and frequency threshold is set to 8. For the Basic Experiments, the PCR is measured separately for each subgroup. Figure 11 shows that high-frequency queries show a notable improvement.

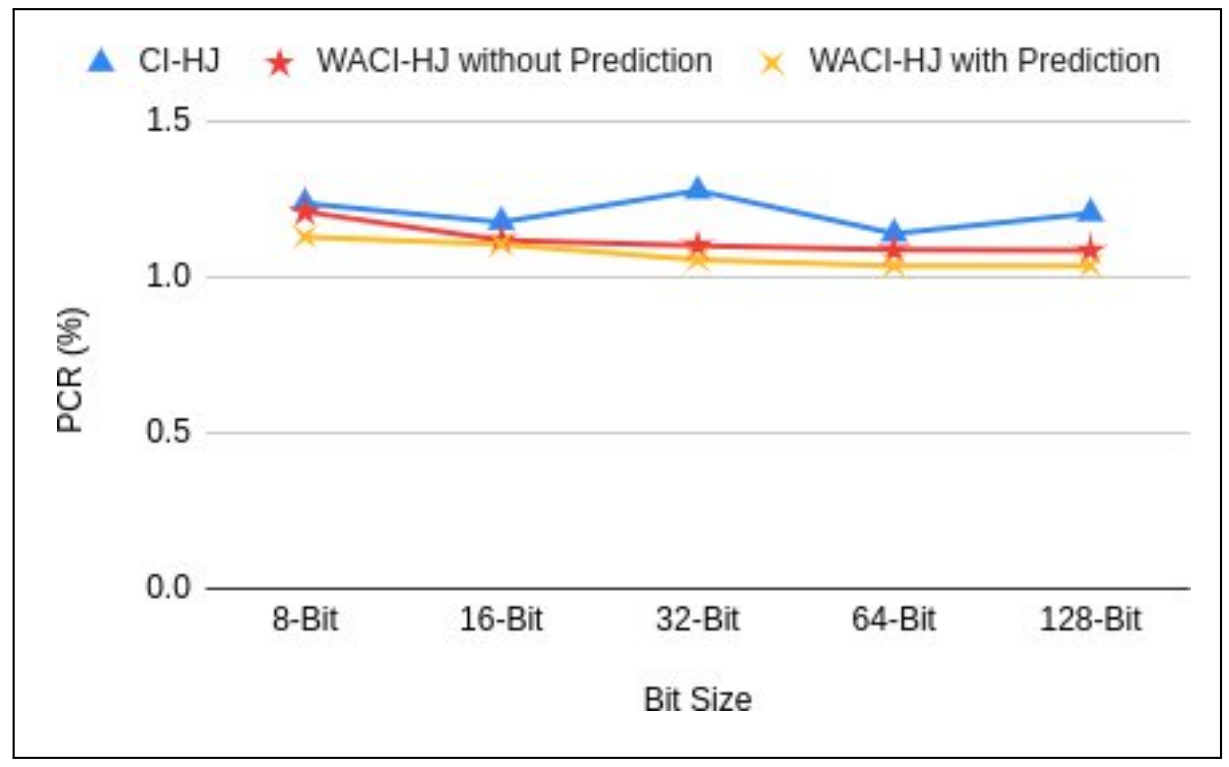

**Figure 7:** Set 0: PCR for Setting Optimal Bit Size

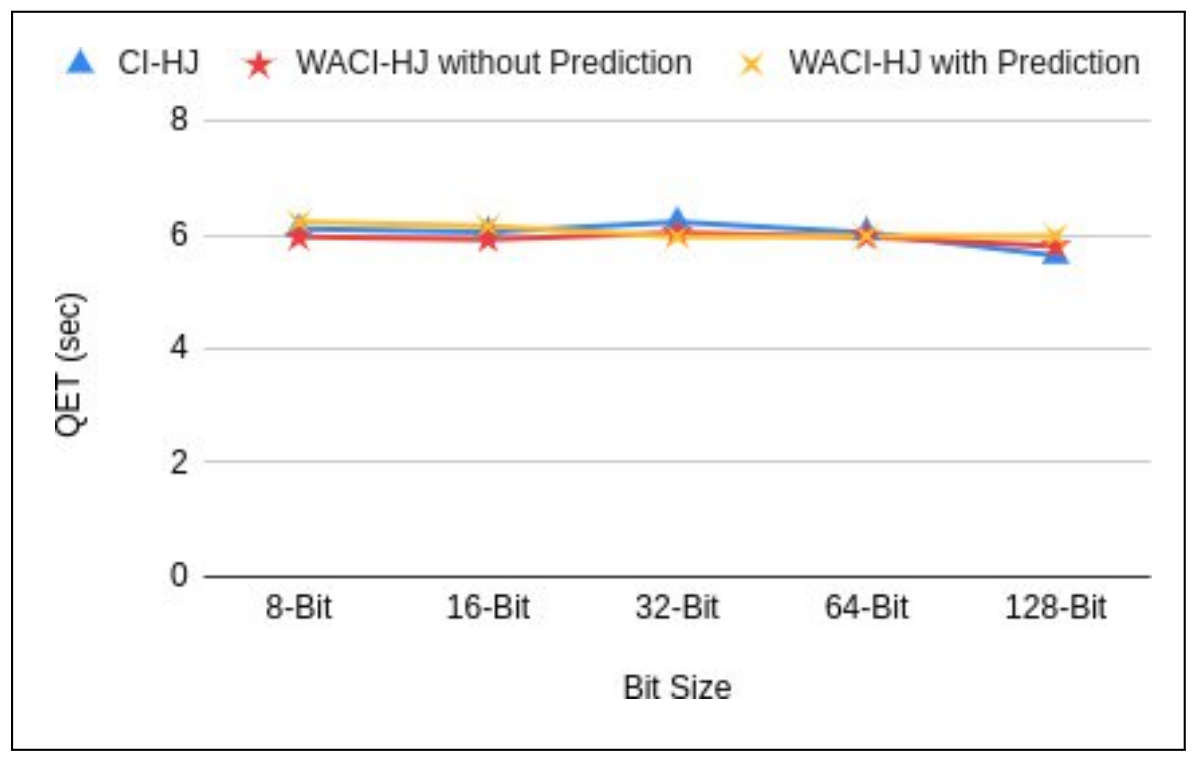

**Figure 8:** Set 0: QET for Setting Optimal Bit Size

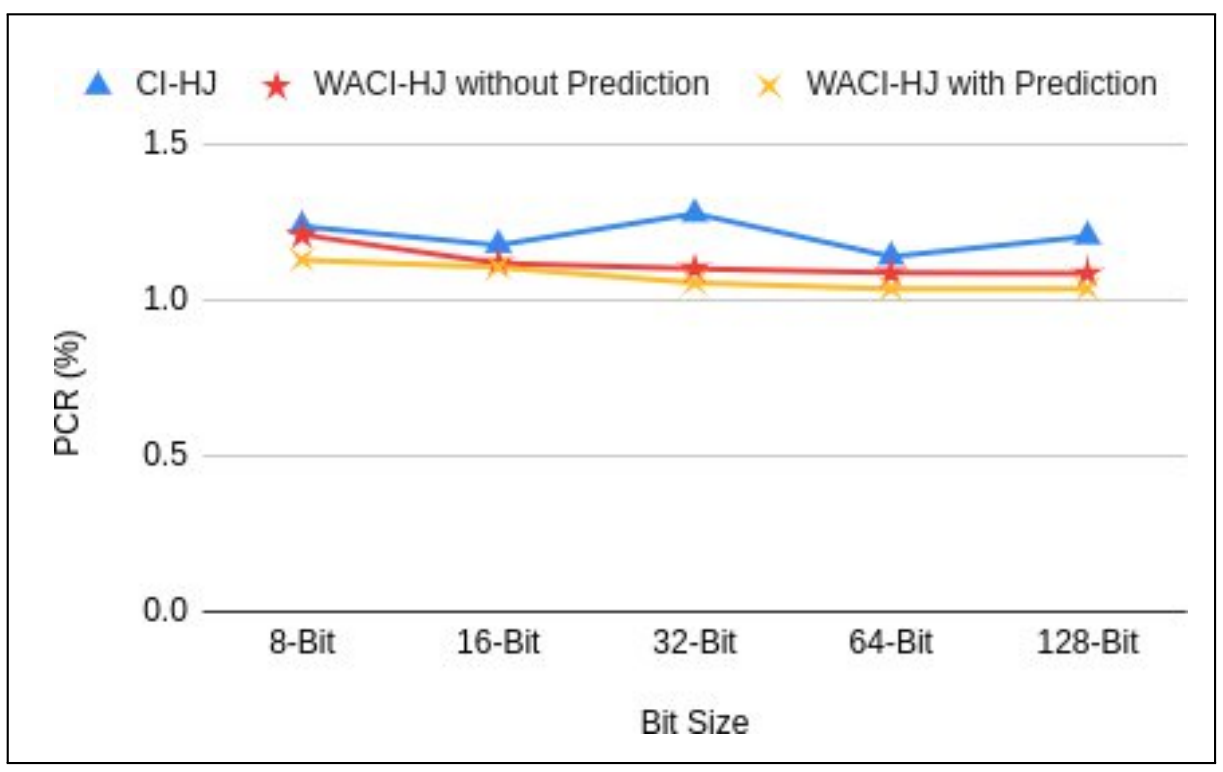

**Figure 9:** MARTA Set 0: PCR for Setting Optimal Number of Clusters

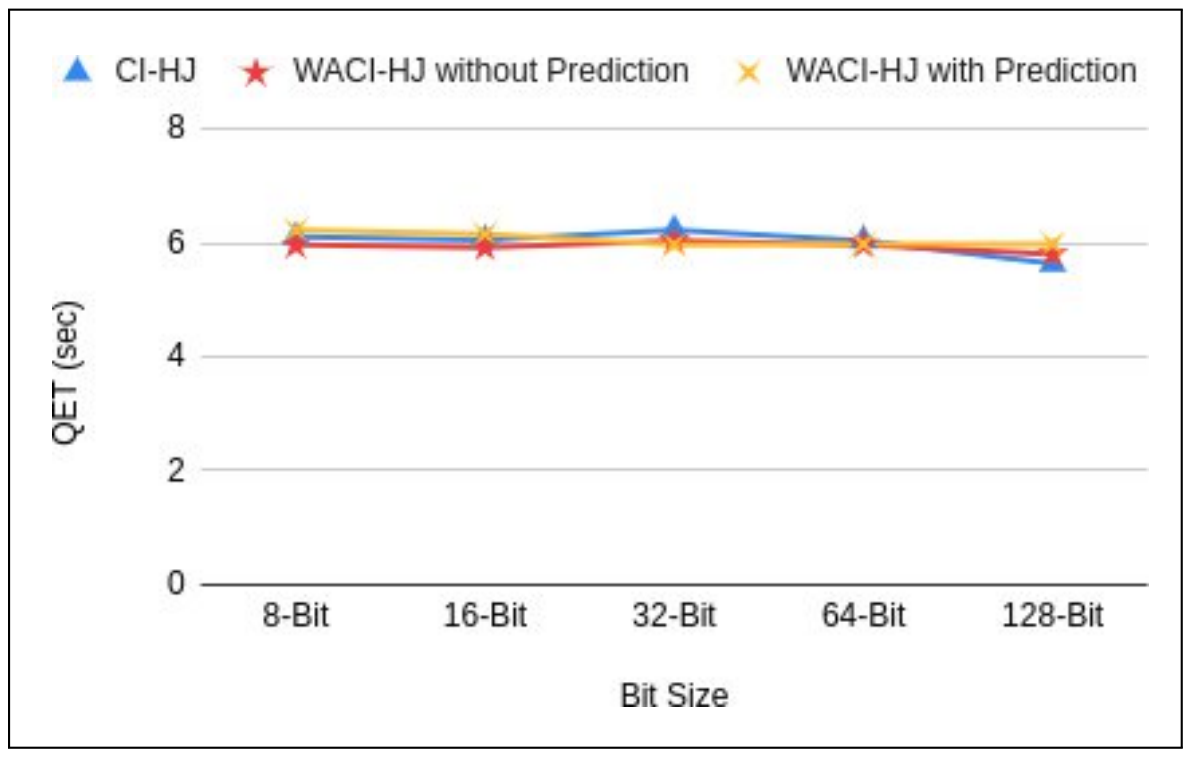

**Figure 10:** MARTA Set 0: QET for Setting Optimal Number of Bit Size

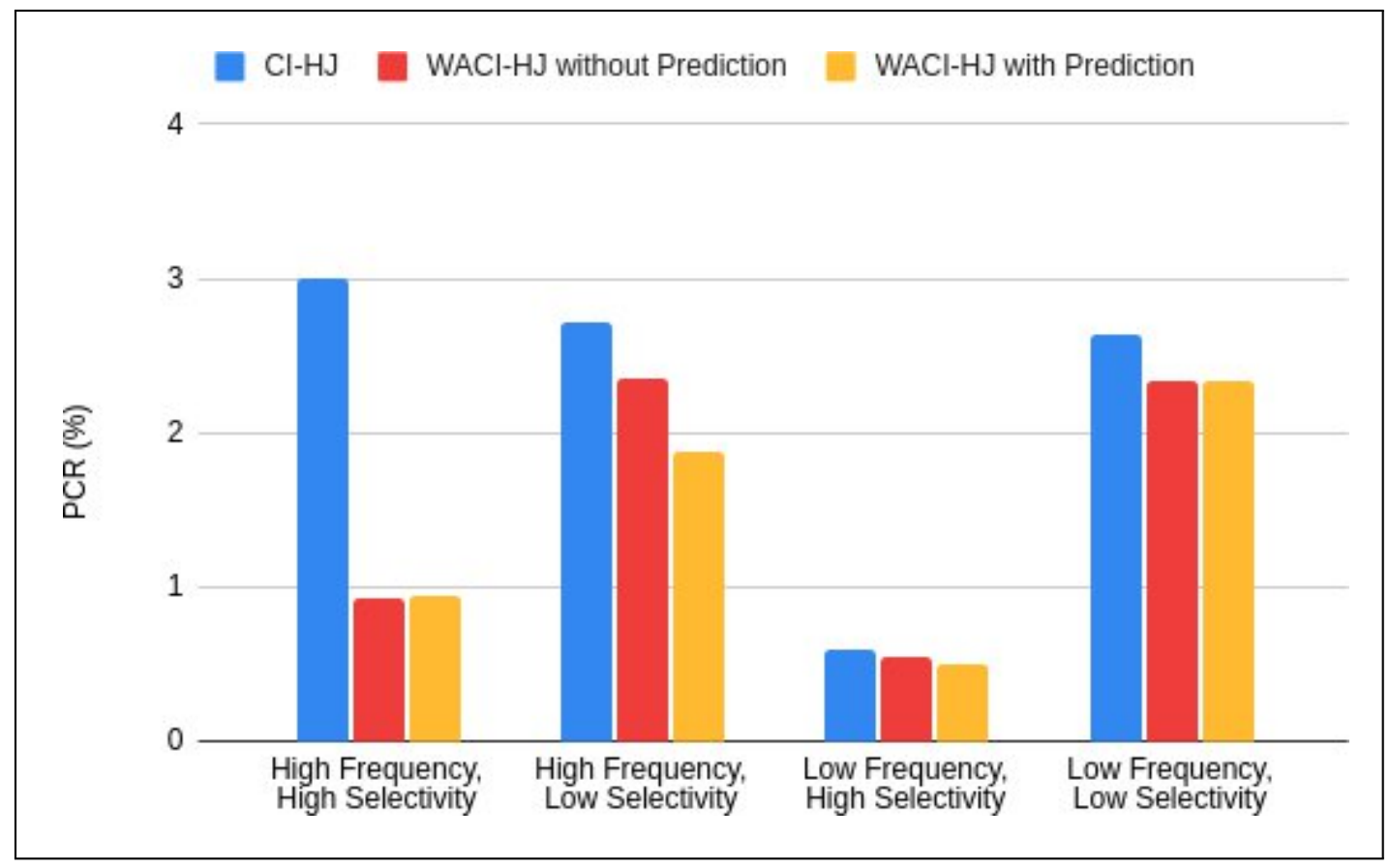

**Figure 11:** MARTA Set I: PCR

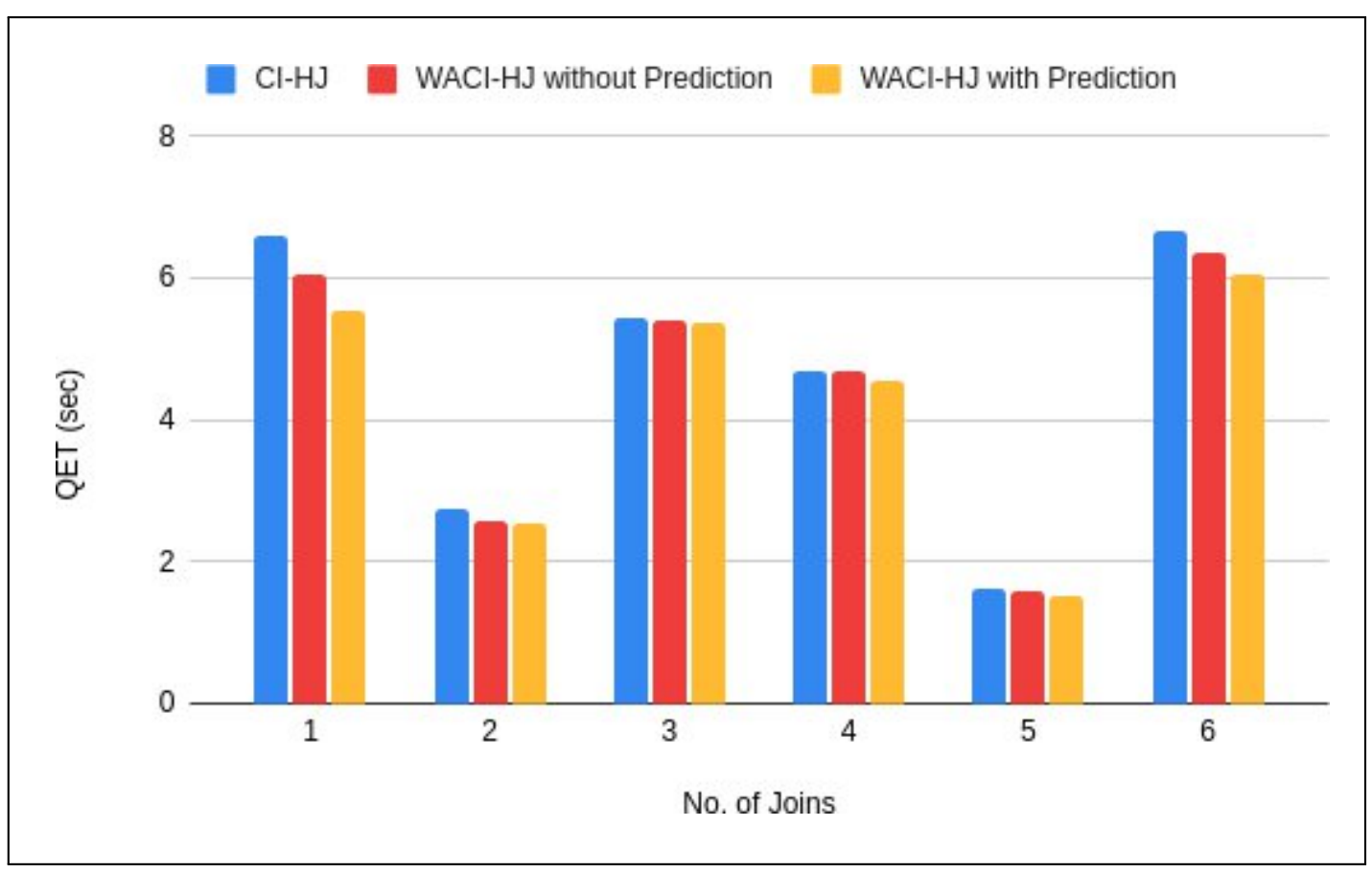

**Figure 12:** MARTA Set I: QET

Using WACI-HJ leads to a 50% decrease in scanned cachelines compared to CI-HJ [14]. Furthermore, integrating workload prediction with WACI-HJ results in an additional 4% improvement. This enhancement is attributed to considering workload information and optimizing based solely on data values. Reducing cacheline scans results in lower RAM and I/O usage, reducing system latency.

**Query Execution Time**

Figure 12 displays the results of the basic QET concerning the number of joins. Since WACI-HJ focuses on accelerating hash joins, queries are categorized based on the number of joins it contains. Improved results are observed for both simple queries with less number of joins and for complex queries with more number of joins, with even greater enhancements for queries involving a higher number of joins. These findings indicate that the WACI-HJ approach is more efficient and effective with join queries than the CI-HJ [14] approach, resulting in faster QET and reduced CPU usage.

The WACI-HJ method achieved a 5% overall improvement compared to the CI-HJ method, with additional enhancements of 5% when workload prediction is integrated with WACI-HJ.

## 5.3. Set II: Data Scaling Experiments

Data Scaling Experiments execute the proposed technique with data sizes ranging from x, 2x, 5x, 7x, and 10x. The results of PCR and QET for the data scaling experiments are shown in this section.

**Percentage of Cachelines Read**

Figure 13 shows the results of the PCR for Data Scaling Experiments. The overall gain was 45% to 51% when scaled from x to 10x. The trail indicates that the PCR for WACI-HJ is constant with respect to data size.

**Query Execution Time**

The results of the scaled QET are depicted in Figure 14. For Data Scaling Experiments, it is observed that the WACI-HJ method achieved an overall improvement of 5% to 10%.

This validates that the observed improvement in QET is consistent even when the scale of the data is increased. These results highlight the potential benefits of using the WACI-HJ method for similar query scenarios, mainly when dealing with large-scale data.

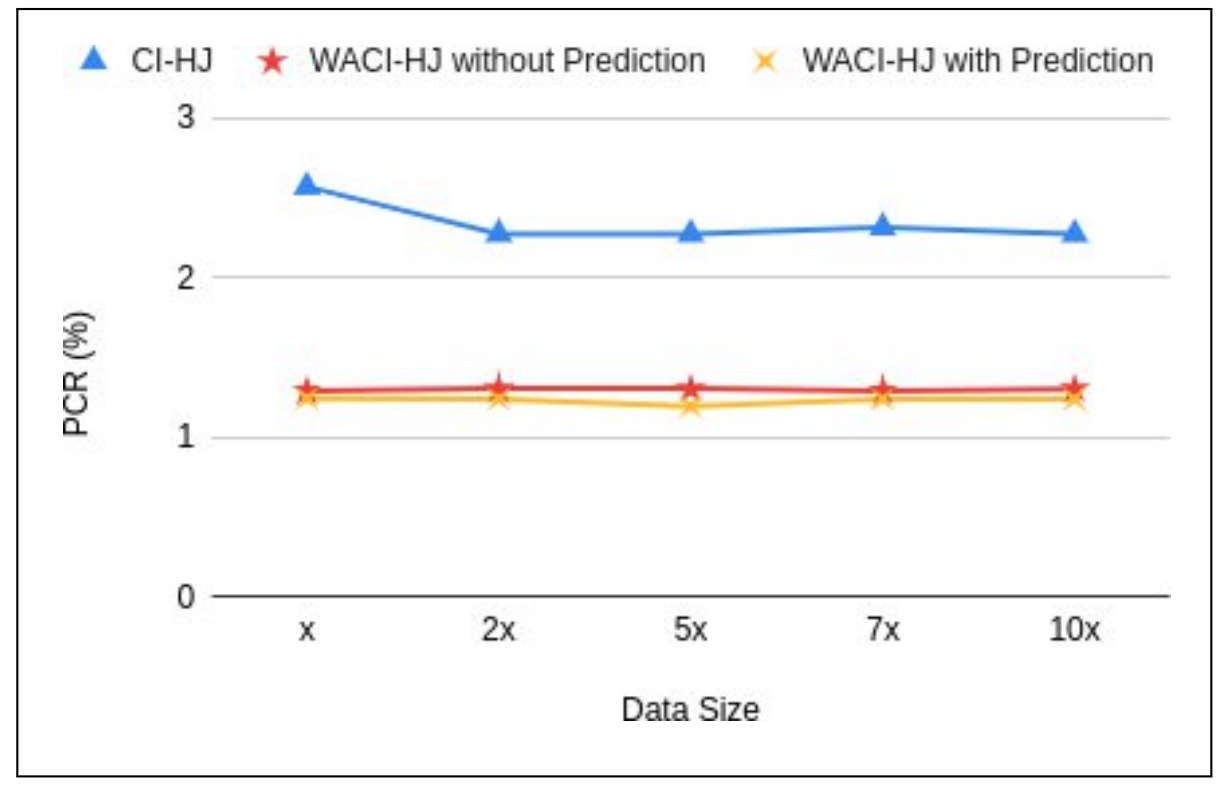

**Figure 13:** MARTA Set II: PCR

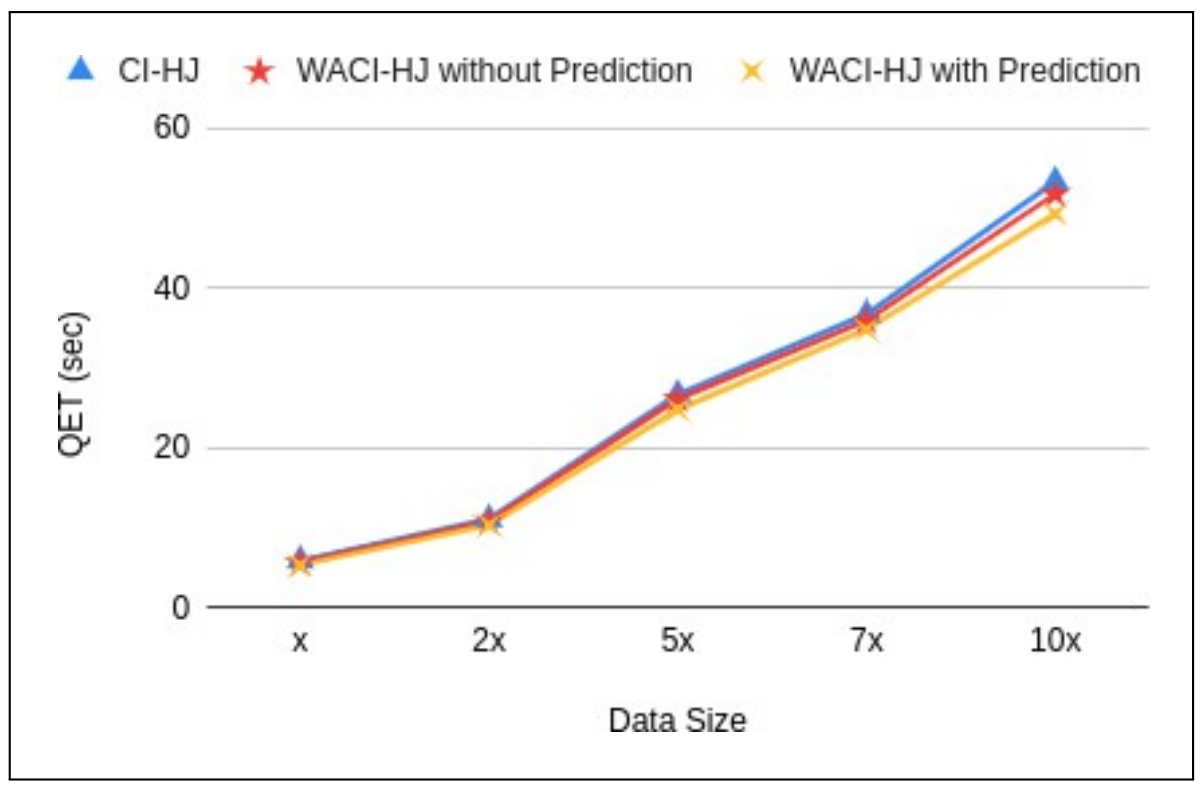

**Figure 14:** MARTA Set II: QET

## 5.4. Set III: Varying Skewness Levels Experiments

Varying Skewness Levels Experiments are performed using the TPC-D [17] dataset. WACI-HJ is evaluated at different levels of data skewness. It is evaluated at 60%, 70%, 80%, 90%, and 99%. 50% is uniformly distributed. The results of PCR and QET for the various skewness levels are depicted in this section.

**Percentage of Cachelines Read**

WACI-HJ PCR is constant for various skewness levels, as shown in Figure 15. An 5% to 50% improvement is recorded for WACI-HJ over CI-HJ as the data gets skewed. A peak is observed at 80% skew, which lies between balanced and extremely skewed distributions. At this level, neither uniform nor skew-aware strategies work well, leading to localized performance issues [61].

**Query Execution Time**

As shown in Figure 16, WACI-HJ becomes 2% to 10% faster with a further increase in skewness level. It shows that WACI-HJ is robust for different skewness levels.

## 5.5. Set IV: Energy-Efficiency Experiments

Although PCR is an indirect indicator of energy consumption, it is directly measured in this section by monitoring CPU, RAM, and I/O resources. PCR for WACI-HJ shows an 49% reduction in energy consumption, making it energy-efficient compared to CI-HJ. WACI-HJ uses fewer resources, which is suitable for a resource-constraint edge system.

The energy-efficiency of WACI-HJ is estimated to be better if Resource Utilization tools are used. Tools such as htop and iotop are used to measure the utilization of CPU, RAM, and I/O. This section shows the resource utilization of the algorithm with respect to the state-of-the-art technique. CPU, RAM, and I/O resource utilization are measured using the MARTA dataset.

**CPU and RAM Utilization**

The CPU utilization of WACI-HJ is shown in Figure 17. A 1% gain is observed in CPU usage compared to CI-HJ, reducing the latency. Figure 18, a 38% gain is observed in RAM usage compared to CI-HJ due to the optimal number of cachelines read.

**I/O Efficiency**

Due to the effective number of cachelines read, I/O operations were reduced by 49%, making the system efficient and faster, as seen in Figure 19. WACI-HJ offers a favourable trade-off between accuracy and resource utilization, making it suitable for resource-constrained edge environments.

WACI-HJ incurs a 2% higher AET than CI-HJ [14] due to delays from generating workload-based bins for optimized joins. Despite this, WACI-HJ improves QET by 10% for MARTA. As data size scales from x to 10x, WACI-HJ achieves a faster AET than CI-HJ, benefiting from preset workload bins that reduce overhead in large scale edge systems.

## 5.6. Comparison with State-of-the-Art

The state-of-the-art CI-HJ [14] uses CIs with hash joins to enhance query performance, as CIs avoid scanning irrelevant data blocks. This method has enhanced the hash join performance to some extent. However, the processing overhead could be significant with a data-aware approach for large datasets, as unnecessary data scanning will add to latency. It also lacks the ability to adjust to varying workloads in real-time.

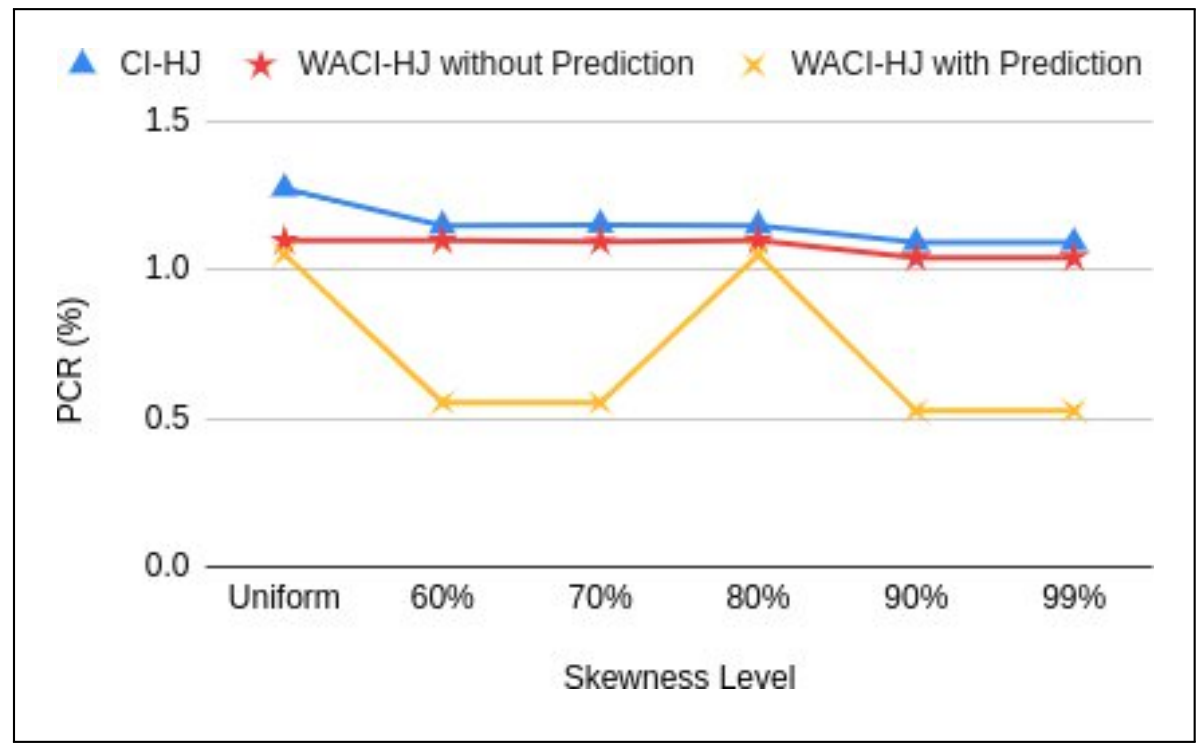

**Figure 15:** TPC-D Set III: PCR

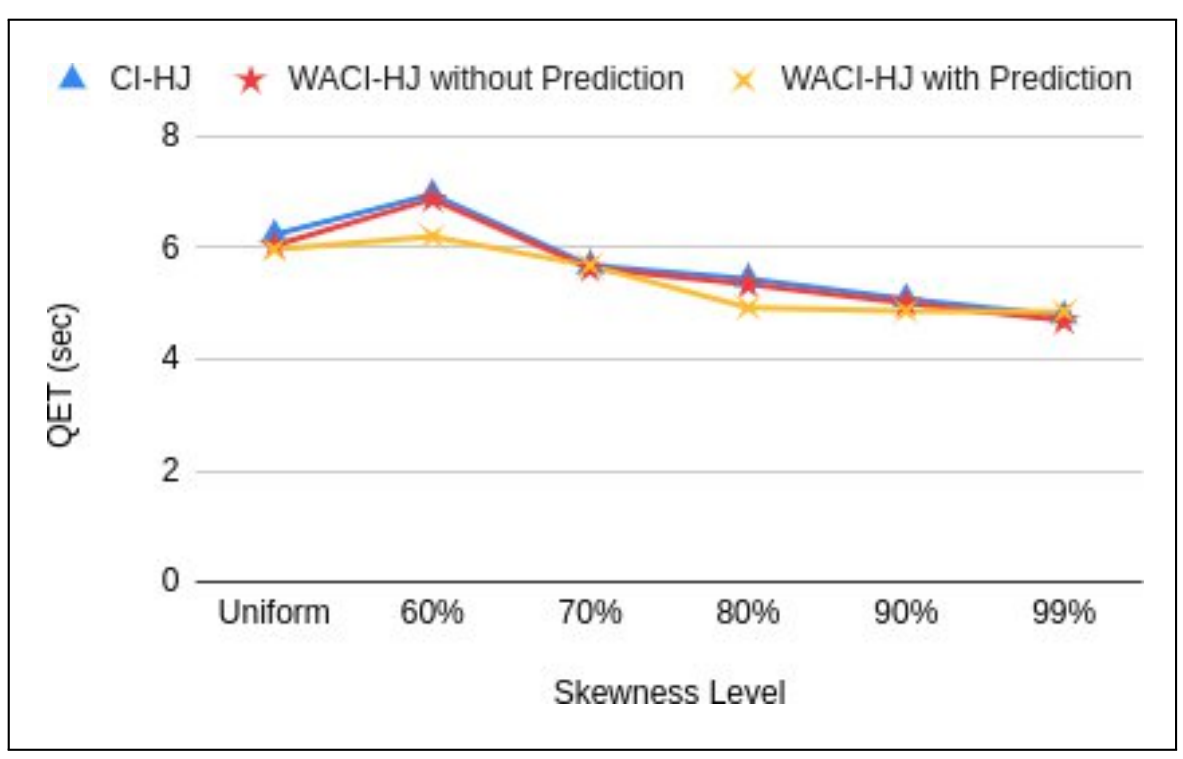

**Figure 16:** TPC-D Set III: QET

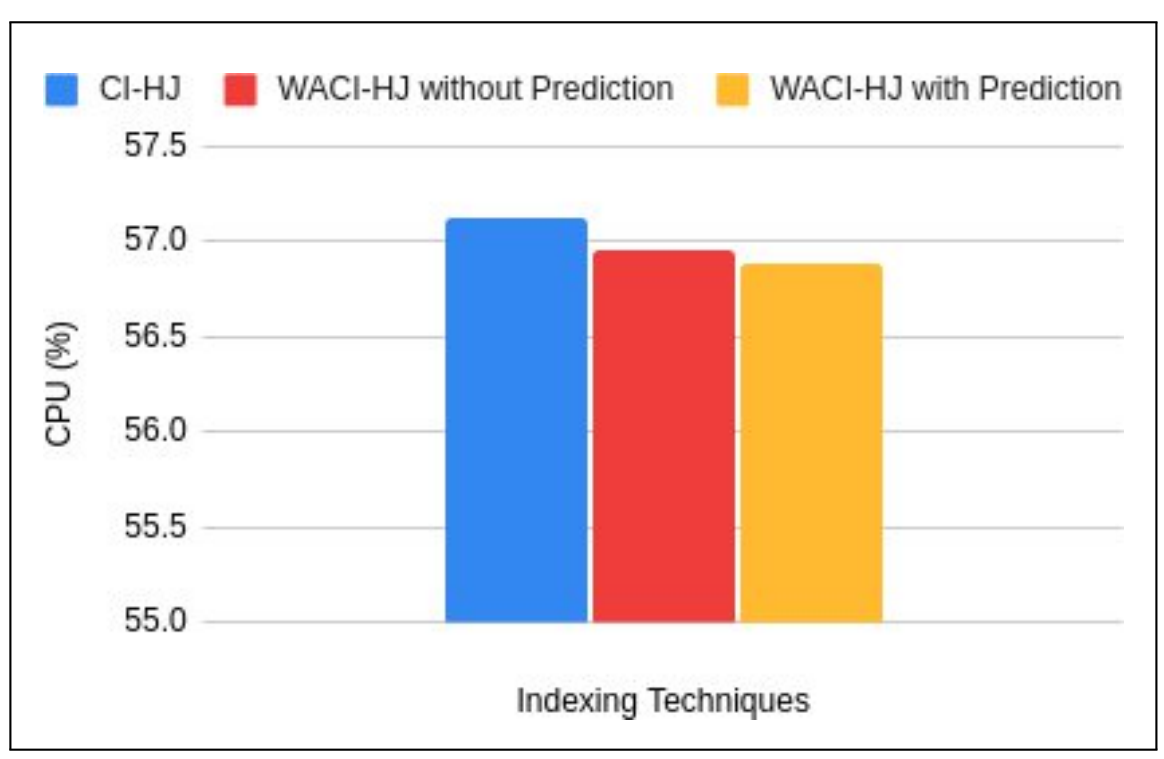

**Figure 17:** MARTA Set IV: CPU Utilization

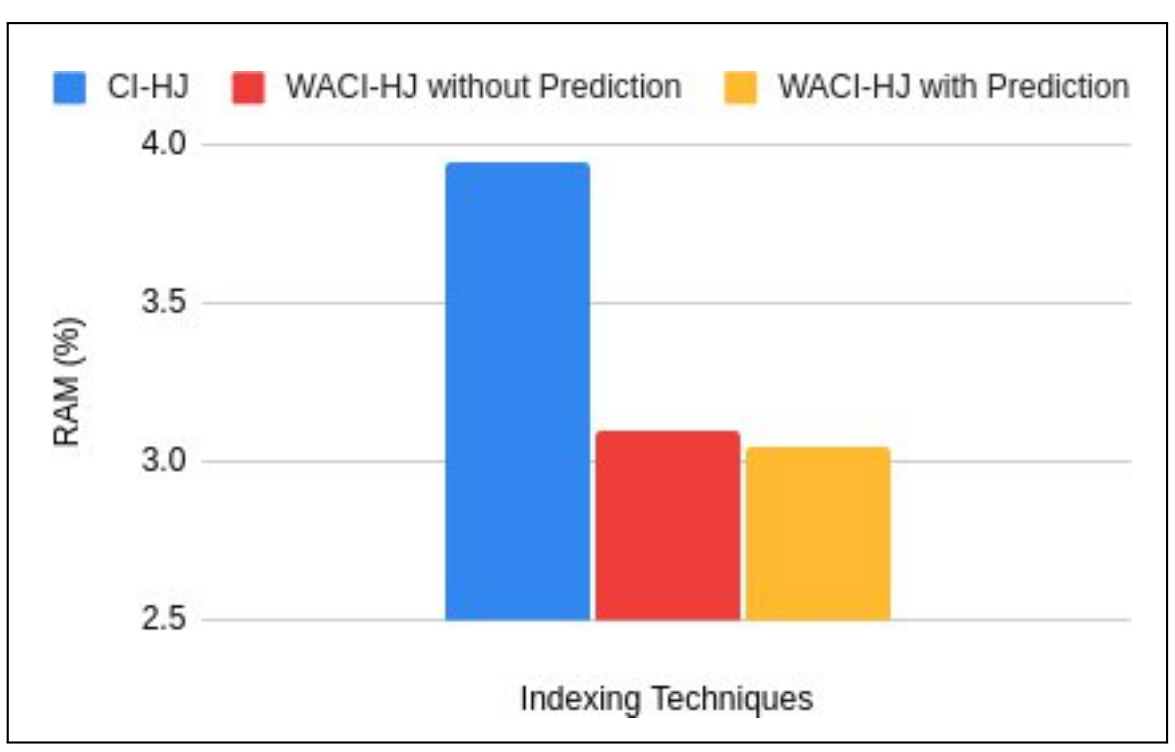

**Figure 18:** MARTA Set IV: RAM Utilization

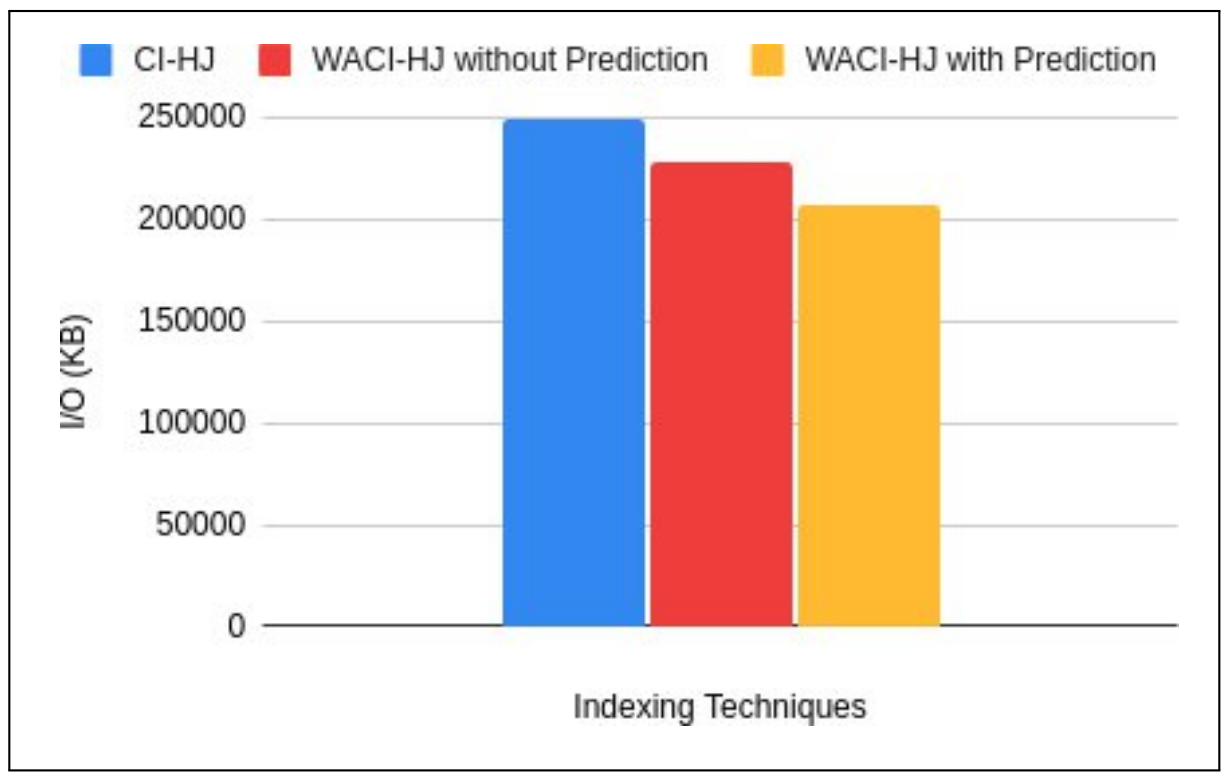

**Figure 19:** MARTA Set IV: IO Efficiency

**Table 2**
Quantitative Comparison

| Parameters/Indexing Techniques | CI-HJ [14] | WACI-HJ | % Gain (%) |
|---|---|---|---|
| **PCR (%)** | 2.57 | 1.24 | 54 |
| **QET (sec)** | 5.97 | 5.38 | 10 |
| **AET (sec)** | 0.25 | 0.28 | -2 |
| **Energy-Efficiency_CPU (%)** | 50.02 | 49.53 | 1 |
| **Energy-Efficiency_RAM (%)** | 3.85 | 2.38 | 38 |
| **Energy-Efficiency_I/O (KB)** | 38174.8 | 19516.8 | 49 |

The proposed technique WACI-HJ uses a workload-aware approach, where the bins are created according to the workload information to reduce cache misses. The upcoming workload is predicted in advance to address real-time query processing.

### *5.6.1. Quantitative Comparison*

Table 2 compares CI-HJ [14] and WACI-HJ quantitatively across various performance metrics. In the context of experiments conducted on the MARTA [18] dataset, WACI-HJ demonstrates significant improvements over CI-HJ. Specifically, WACI-HJ achieves an 54% enhancement in PCR, indicating more efficient query processing than CI-HJ. Moreover, WACI-HJ reduces QET by 10 seconds, showcasing its ability to execute queries more swiftly. Despite a slight increase of 2% in AET, attributed to the prediction overhead for anticipating query workloads, these results underscore WACI-HJ effectiveness in optimizing query performance for MARTA.

The Energy-Efficiency comparison between CI-HJ and WACI-HJ was based on the resources utilized. WACI-HJ demonstrates a notable improvement in RAM utilization efficiency with a gain of 38%, indicating it consumes significantly less memory than CI-HJ. This efficiency is crucial for reducing resource overhead in computational environments, especially those with limited RAM availability. Additionally, WACI-HJ achieves an 49% gain in I/O operations, highlighting its capability to minimize data access requirements, which is beneficial for optimizing overall system performance and energy consumption. However, regarding CPU utilization, WACI-HJ shows only a marginal 1% gain over CI-HJ, suggesting similar efficiency levels in CPU resource management between the two indexing techniques. These findings underscore WACI-HJ effectiveness in enhancing resource utilization and performance, particularly in resource-constraint edge environments.

### *5.6.2. Qualitative Comparison*

Table 3 provides a Qualitative Comparison between two indexing techniques, CI-HJ [14] and WACI-HJ, highlighting its capabilities and performance across several key parameters. CI-HJ is a data-aware approach capable of handling scaled and skewed data, but it scans unnecessary cachelines and lacks real-time query processing. Additionally, it does not include predictive features and is not energy-efficient. CI-HJ performs better on synthetic datasets but falls short on benchmark datasets like TPC-H, indicating limited applicability in real-world scenarios.

In contrast, WACI-HJ exhibits several advantages over CI-HJ [14] as it is a workload-aware approach that avoids scanning unnecessary cachelines and can handle real-time query processing. It excels at managing extensively skewed data to capture variations at the minute points. It includes predictive capabilities and optimizing performance based on the data access patterns. Furthermore, WACI-HJ is considered more energy-efficient, making it a sustainable choice for large-scale applications. Its performance on benchmark datasets like TPC-H and real-world datasets like MARTA demonstrates its robustness and versatility, making WACI-HJ a more effective and broadly applicable indexing technique for diverse database environments.

## 5.7. WACI-HJ Performance for Different Datasets

The experiments were performed across the set of experiments using a benchmark dataset and a real-world dataset. Benchmark TPC-H [16] and TPC-D [17] belong to the supply chain domain of EC applications. Additionally, MARTA [18] represents a real-world dataset in transportation, widely utilized across numerous EC applications. TPC-H and MARTA %gain with respect to technique state-of-the-art CI-HJ [14] is depicted in Figure 20. The listed parameters are PCR, QET, and AET.

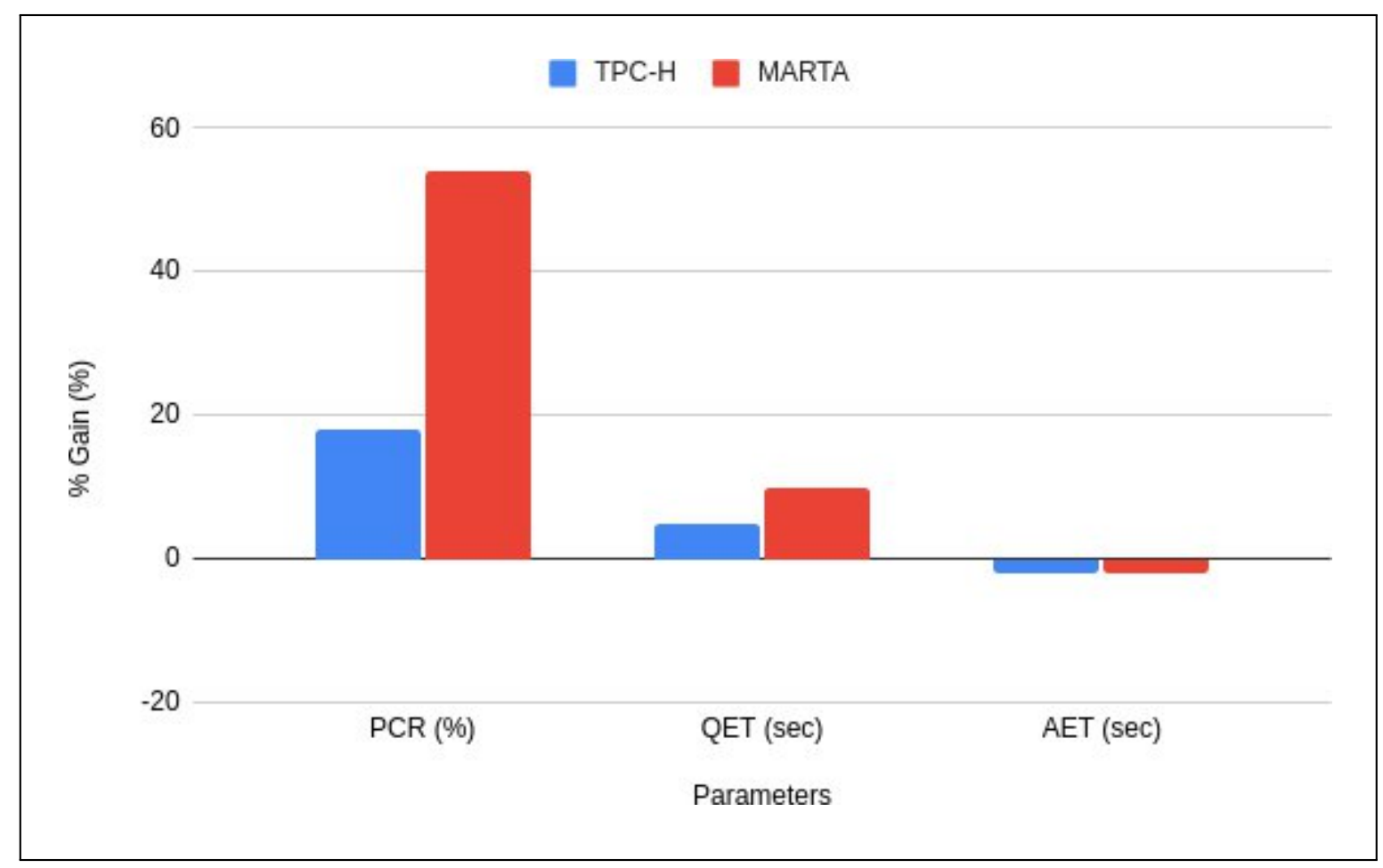

**Figure 20:** TPC-H vs. MARTA Comparison

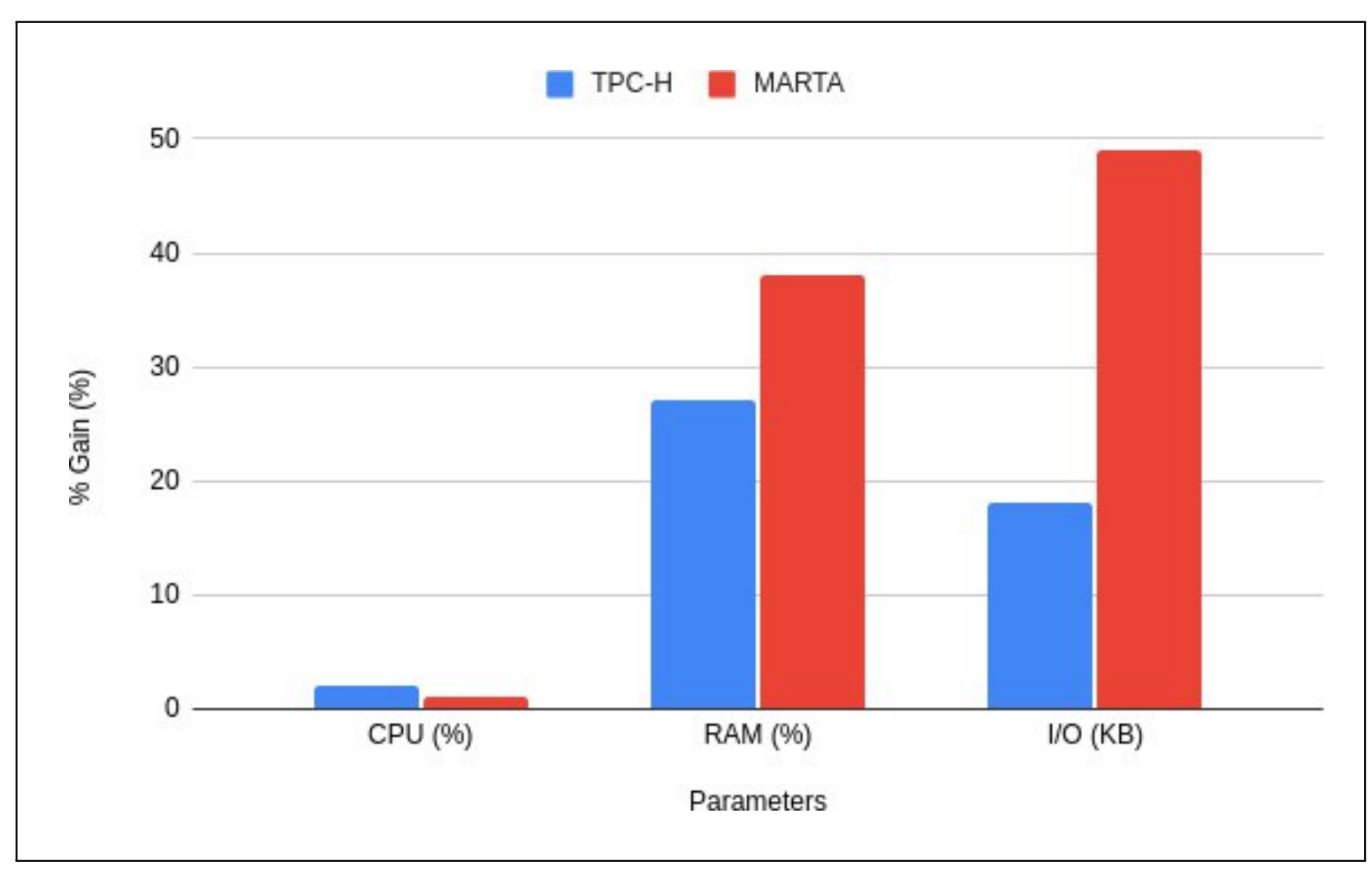

**Figure 21:** TPC-H vs. MARTA Energy-Efficiency Comparison

PCR indicates significant improvements of 18% and 54% for TPC-H and MARTA, respectively. TPC-H shows notable enhancement in lower-selectivity queries due to a higher proportion of non-distinct values, whereas MARTA demonstrates its greatest improvement in frequent queries. Furthermore, using a workload-aware approach in these experiments results in a 5% and 10% overall improvement for TPC-H and MARTA, respectively, with larger gains observed in queries involving multiple joins. AET takes 2% overhead for both datasets, as it needs to predict the query workload. But, For larger data sizes, WACI-HJ performs better than CI-HJ for both TPC-H and MARTA, which is evident in AET.

**Table 3**
TPC-H vs. MARTA Quantitative Comparison

| **Parameters/% Gain** | **TPC-H [16] Gain (%)** | **MARTA [18] % Gain (%)** |
|---|---|---|
| **PCR (%)** | 18 | 54 |
| **QET (sec)** | 5 | 10 |
| **AET (sec)** | -2 | -2 |
| **Energy-Efficiency_CPU (%)** | 2 | 1 |
| **Energy-Efficiency_RAM (%)** | 27 | 38 |
| **Energy-Efficiency_I/O (KB)** | 18 | 49 |

**Table 4**
Qualitative Comparison

| Indexing Techniques/Parameters | Workload-Aware | Scaled Data | Skewed Data | Prediction | Energy-Efficient | Performance over Different Datasets |
|---|---|---|---|---|---|---|
| **CI-HJ [14]** | No | Yes | Yes | No | Less | Improved results obtained for Synthetic Dataset, but not on Benchmark Dataset TPC-H |
| **WACI-HJ** | Yes | Yes | Extensively Skewed Data | Yes | More | Better results obtained for both Benchmark Datasets TPC-H and a Real-World Dataset MARTA |

The energy-efficiency comparison between TPC-H and MARTA is depicted in Figure 21. WACI-HJ proves more efficient by requiring 18% and 49% fewer I/O scans than CI-HJ for TPC-H and MARTA, respectively. This efficiency is particularly advantageous in resource-constrained edge environments. These results are summarized in form of a table in Table 4.

# 6. Conclusions and Future Work

The WACI-HJ technique addresses challenges in managing and processing data at the edge, where resource constraints are prevalent. It improves upon the CI-HJ method by employing workload-aware strategies and predictive data storage optimization, enhancing algorithm performance before queries arrive.

The experimental results demonstrate that WACI-HJ significantly reduces the PCR and QET compared to CI-HJ. Demonstrating with a real-world Smart Transportation dataset validates the effectiveness of WACI-HJ, with a 54% improvement in PCR and a 10% improvement in QET. Although PCR is an indirect measure of energy consumption, while energy-efficiency experiments measure energy consumption directly, the notable gains are observed in resource utilization, i.e. 1% in CPU, 38% in RAM, and 49% in I/O.

In edge query processing, WACI-HJ significantly reduces query latency and enhances energy-efficiency compared to the state-of-the-art CI-HJ. By optimizing data access patterns and incorporating predictive capabilities, WACI-HJ minimizes query response times and conserves energy resources. By minimizing cacheline reads and optimizing query execution, the solution enhances real-time data processing in resource-constrained edge environments, enabling faster traffic analysis, congestion management, and intelligent routing in smart transportation systems. Additionally, this technology can be applied to other domains to accelerate edge query processing.

Several future directions have been identified, including dynamic resource allocation [62], generating multi-attribute column imprints [63], and integrating RDF data types [64]. These aspects aim to enhance system efficiency, improve data characterization, and explore new data representation formats for advanced query processing and optimization.